\documentclass[12pt]{article}
\usepackage{graphicx, amssymb, amsthm, amsbsy, amsmath,bm}
\usepackage{color,xcolor}
\usepackage{url}
\usepackage[colorlinks=true,allcolors=blue]{hyperref}%
\usepackage{algorithm}
\usepackage{multirow}
\usepackage{algorithmic}
\usepackage{natbib}
\usepackage{dblfloatfix}
\usepackage{epsfig}
\usepackage[export]{adjustbox}
\usepackage{subfigure}
\usepackage[margin=1in]{geometry}

\DeclareGraphicsExtensions{.pdf,.eps.gz,.eps,.epsi.gz,.epsi,.ps,.ps.gz}
\long\def\symbolfootnote[#1]#2{\begingroup%
\def\thefootnote{\fnsymbol{footnote}}\footnote[#1]{#2}\endgroup}

\newcommand{\Var}{\mathrm{Var}}
\newcommand{\Cov}{\mathrm{Cov}}

\newcommand{\I}{{\bm I}}

\newcommand{\bSigma}{{\bm \Sigma}}
\newcommand{\bW}{{\bm W}}
\newcommand{\bpi}{{\bm \pi}}

\newcommand{\blambda}{{\bm \lambda}}
\newcommand{\bOmega}{{\bm \Omega}}

\def\A{{\bm A}}
\def\a{{\bm a}}
\def\B{{\bm B}}
\def\c{{\bm c}}
\def\D{{\bm D}}
\def\G{{\bm G}}
\def\V{{\bm V}}

\def\H{{\bm H}}

\def\b{{\bm b}}
\def\I{{\bm I}}

\def\T{{\bm T}}

\def\U{{\bm U}}

\def\u{{\bm u}}
\def\v{{\bm v}}
\def\W{{\bm W}}
\def\w{{\bm w}}
\def\X{{\bm X}}
\def\x{{\bm x}}

\def\y{{\bm y}}

\def\z{{\bm z}}
\def\1{{\bf 1}}
\def\0{{\bf 0}}
\def\R{{\bm R}}
\def\e{{\bm e}}
\def\M{{\bm M}}

\newcommand\hzc[1]{{\color{cyan}{[HZ: #1]}}}

\def\cb{\textcolor{orange}}

\newtheorem{thm}{Theorem}%
\newtheorem{lemma}{Lemma}%
\newtheorem{prop}{Proposition}

\title{\bf Statistical Inference of Cell-type Proportions Estimated from Bulk Expression Data}
\author{
Biao Cai$^\text{a}$, Jingfei Zhang$^\text{b}$, Hongyu Li$^\text{a}$, Chang Su$^\text{a}$ and Hongyu Zhao$^\text{a}$
\medskip \\
$^\text{a}$ {\normalsize Department of Biostatistics, Yale University}\\ 
$^\text{b}$ {\normalsize Department of Management Science, University of Miami}\\
}
\date{}

\begin{document}

\maketitle
\renewcommand{\baselinestretch}{1.35}
\begin{abstract}

\noindent
There is a growing interest in cell-type-specific analysis from bulk samples with a mixture of different cell types. A critical first step in such analyses is the accurate estimation of cell-type proportions in a bulk sample. Although many methods have been proposed recently, quantifying the uncertainties associated with the estimated cell-type proportions has not been well studied. Lack of consideration of these uncertainties can lead to missed or false findings in downstream analyses. In this article, we introduce a flexible statistical deconvolution framework that allows a general and subject-specific covariance of bulk gene expressions. Under this framework, we propose a decorrelated constrained least squares method called DECALS that estimates cell-type proportions as well as the sampling distribution of the estimates. Simulation studies demonstrate that DECALS can accurately quantify the uncertainties in the estimated proportions whereas other methods fail. Applying DECALS to analyze bulk gene expression data of post mortem brain samples from the ROSMAP and GTEx projects, we show that taking into account the uncertainties in the estimated cell-type proportions can lead to more accurate identifications of cell-type-specific differentially expressed genes and transcripts between different subject groups, such as between Alzheimer's disease patients and controls and between males and females.
\end{abstract}


\newpage
\baselineskip=26.5pt
\section{Introduction}
The need to analyze gene expression data collected either through microarrays or sequencing to answer different biological questions has motivated the developments of a great number of statistical methods in the last two decades. The applications of these methods have gained novel biological insights on disease mechanisms, identified informative biomarkers, and led to novel treatments for some diseases \citep[e.g.,][]{zhang2005general, barabasi2011network, trapnell2012differential, zhang2013integrated, mostafavi2018molecular, zhang2022high}. 
While the literature on gene expression analysis is steadily growing, most gene expression data gathered to date are from bulk samples which consist of distinct cell types. For example, a brain sample usually has astrocytes, endothelial cells, microglia, neurons, oligodendrocytes, and oligodendrocyte precursor cells \citep{darmanis2015survey}. Therefore, even if two samples have the same gene expression profiles at the cell type level, their aggregated bulk expression profiles may differ if their cell-type proportions are different. 
Due to the heterogeneous cell-type proportions across samples, the analysis of gene expression data at the bulk level may lead to false positive findings and miss true biological signals. 
Moreover, such an analysis only offers an aggregated view of the biological mechanisms in different cell types, while most disease etiologies are cell-type-specific \citep{hekselman2020mechanisms}.
To gain a more accurate and comprehensive view of the underlying biological mechanisms, a desirable approach is to analyze gene expressions in specific cell types.
While such cell-type-specific gene expressions are not directly available from bulk sample data, it can be inferred if the cell-type proportions for bulk samples are given. This task of inferring cell-type-specific proportions and/or expressions from bulk samples is often referred as \textit{deconvolution}.

In recent years, many deconvolution methods have been proposed \citep{abbas2009deconvolution,newman2015robust,wang2019bulk,jew2020accurate,tang2020nitumid,yang2021adroit}. These methods rely on the availability of signature genes for different cell types with their expressions usually gathered from single-cell RNA sequencing (scRNA-seq) data, and they differ in the details on how the information from these signature genes is utilized.
The estimated cell-type proportions from these methods together with the bulk expression data have made it possible to address a number of important cell-type-specific (CTS) biological questions.
For example, based on gene expression data collected from two groups of bulk samples, it may be possible to infer 
genes having different CTS expression levels between the two groups 
\citep{jin2021cell,wang2021bayesian,tang2022scadie}. It is also possible to infer CTS co-expression patterns \citep{su2021csnet} and CTS expression quantitative trait loci (eQTLs) \citep{patel2021cell,little2022cell} from bulk samples. Furthermore, instead of making group-level CTS inference, methods have also been proposed to infer CTS expression levels at the individual sample level \citep{newman2019determining,jaakkola2022estimating}. These sample-level inferred CTS expressions have been used to infer CTS differentially expressed genes between groups, genetic variants that have CTS effects on gene expressions, and CTS co-expressions \citep{jin2021cell,wang2021bayesian,jaakkola2022estimating}. 

The majority of the above CTS analysis methods implicitly assume that the true cell-type proportions for bulk samples are available, even though they are often estimated with errors from deconvolution models. 
Limited efforts have been made to investigate and quantify the impacts of uncertainties in estimated cell-type proportions on downstream CTS analysis methods, even though not considering such uncertainties in estimated cell-type proportions
can lead to missed or false findings in downstream CTS analyses.
Two recent methods have been proposed in the literature to quantify the uncertainties in estimated cell-type proportions. \cite{erdmann2021likelihood} proposed a likelihood-based deconvolution method using single-cell reference data, referred to as \texttt{RNA-Sieve}, and confidence intervals of the estimated proportions can be calculated as a by-product of the estimation procedure. Their approach assumes that the error terms from modeling the signature gene expressions are independent and Gaussian. However, this assumption will likely fail for real data because there are correlations among genes and RNA-seq data are more appropriately modeled by non-normal distributions, e.g. negative binomial distributions. 
\cite{xie2022robust} developed a method based on a measurement error model that incorporates the errors in inferring signature gene expression levels from single cell data in the estimates of cell-type proportions in bulk samples, referred to as \texttt{MEAD}. 
The estimated proportions are shown to be asymptotically normal and the covariance is estimated through a sandwich type estimator with an estimated gene-gene dependence set. 
However, this covariance estimator may be biased as the subject-specific covariance among signature genes is not consistently estimated in \texttt{MEAD}, and this can reduce the accuracy of inferential tasks such as constructing confidence intervals. 
Specifically, in our simulation studies in Section \ref{set2}, we show that the confidence intervals for cell-type proportions calculated using \texttt{RNA-Sieve} and \texttt{MEAD} both suffer from under-coverage, sometimes substantially. 

In this paper, we 
develop a statistical deconvolution framework to estimate the cell-type proportions and their sampling distributions, and to incorporate the uncertainties in downstream CTS analysis methods.  
Our approach does not impose parametric assumptions on the distributions of bulk expressions and allows a general covariance among the signature genes that can be cell-type- and subject- specific. 
Specifically, we consider a decorrelated constrained least squares framework (\texttt{DECALS}) to estimate the cell-type proportions, such that the estimated proportions are non-negative and add up to 1, and the distribution of the estimated proportions is derived by decorrelating the signature gene expressions in each bulk sample via their sample-specific covariance.
One major challenge in estimating the distribution of estimated proportions in a sample, say $i$ denoted as $\bpi_i$, is the need to characterize the covariance among signature gene expressions in this sample, denoted as $\bSigma_i$. As bulk expressions are aggregated over different cell types, covariance $\bSigma_i$ is a function of $\bpi_i$ and the unknown CTS covariances. 
To consistently estimate the CTS covariances, we consider a novel moment-based estimator that borrows information across all bulk samples and further consider a finite sample bias correction to improve accuracy. We demonstrate in simulation studies that \texttt{DECALS} can accurately quantify the uncertainties in the estimated proportions whereas other methods fail to offer accurate uncertainty estimates. In Section \ref{sec::real}, we apply \texttt{DECALS} to analyze bulk gene expression data from post mortem brain samples from the ROSMAP and GTEx projects and show that taking into account the uncertainties in the estimated cell-type proportions can lead to more accurate identifications of cell-type-specific differentially expressed genes and transcripts between different subject groups, such as between Alzheimer's disease patients and controls and between males and females. As \texttt{DECALS} is flexible, easy to compute and free from parametric assumptions, it can be easily combined with most CTS analysis methods based on bulk samples to incorporate the uncertainties of estimated cell-type proportions and improve the accuracy and interpretability of the biological findings.

The rest of the paper is organized as follows. 
Section \ref{sec::sec2} introduces the cell type convolution model, the estimation of cell-type proportions and their sampling distributions. 
Section \ref{sec::sim} reports the simulation results. Section \ref{sec::real} performs downstream analysis to identify CTS differentially expressed genes and transcripts between groups of samples for two real studies, demonstrating that taking into account the uncertainties in the cell-type proportions can lead to more enriched and interpretable biological findings.
The paper is concluded with a discussion section.

\section{Estimation and Inference of cell-type proportions}\label{sec::sec2}
\subsection{Cell type deconvolution model}\label{sec::model}
Suppose we have gene expression data $\y_1,\ldots,\y_n \in\mathbb{R}^{p}$ collected from $n$ bulk RNA-seq samples across $p$ signature genes.
We assume that there are $K$ cell types, and the bulk level expression for sample $i$ is the sum of these $K$ cell types written as 
\begin{equation}\label{eq:decomp}
\y_i=\sum_{k=1}^K\pi_{ik}\x_i^{(k)},
\end{equation}
where $\pi_{ik}$ and $\x_i^{(k)}$ are the proportion and expression profile of cell type $k$ in sample $i$, respectively, and $\sum_{k=1}^K\pi_{ik}=1$. 
In this paper, we do not make any parametric assumptions on the distributions of CTS expression profile $\x_i^{(k)}$ and bulk expression $\y_i$. 
Denoting $\mathbb{E}(\x_i^{(k)})=\w_k$, where $\w_k$ represents the signature gene expression profile for the $k$th cell type, we may write
\begin{equation}\label{eqn:de}
\y_i=\sum_{k=1}^K\pi_{ik}\w_k+\bm{\epsilon}_i,\quad \mathbb{E}({\bm\epsilon}_i)=0,
\end{equation}
where 
$\bm\epsilon_i=(\epsilon_{i1},\ldots, \epsilon_{ip})$ is a vector of random variables with mean zero.
In cell type deconvolution analysis, the CTS mean expressions $\{\w_k\}_{1\le i\le K}$ are usually gathered from pure cell types \citep{newman2015robust, li2016comprehensive} or scRNA-seq data \citep{wang2019bulk, newman2019determining, jew2020accurate}. 
Given bulk expressions $\{\y_i\}_{1\le i\le n}$ and CTS mean expressions $\{\w_k\}_{1\le i\le K}$, we focus on the inference of $\{\bpi_i\}_{1\le i\le n}$, where $\bpi_i=(\pi_{i1},\ldots,\pi_{iK})$ denotes the vector of cell-type proportions in sample $i$.

Before we proceed, we first highlight some important differences between \eqref{eqn:de} and a standard linear regression problem.
{\it First}, model \eqref{eqn:de} estimates $\bpi_i$ with $p$ observations $(y_{i1},\ldots,y_{ip})$ representing the bulk expressions of $p$ signature genes in sample $i$.
The statistical units in \eqref{eqn:de} are the $p$ signature genes, rather than the $n$ bulk samples. 
Hence, the estimation accuracy of $\bpi_i$ is expected to be more closely related to $p$, the number of signature genes, than $n$, the number of samples.
{\it Second}, the error terms $(\epsilon_{i1},\ldots, \epsilon_{ip})$ in \eqref{eqn:de} are not independent. Specifically, $\Cov(\bm\epsilon_i)$ can be written as a sum of CTS covariances between the signature genes weighted by cell-type proportions $(\pi_{i1},\ldots,\pi_{iK})$; see \eqref{ass1eq}.
As a result, drawing inference on $\bpi_{i}$ via \eqref{eqn:de} demands estimating $\Cov(\bm\epsilon_i)$, termed {\it subject-specific covariance} in this paper.
{\it Third}, as $\pi_{ik}$'s are cell-type proportions in sample $i$, they must satisfy the constraints that $\pi_{ik}\geq 0$ and $\sum_{k=1}^K\pi_{ik}=1$. 
The above unique aspects in \eqref{eqn:de} pose new and significant challenges in the statistical inference of cell-type proportions, which we will address in the ensuing development.

\subsection{Estimation of cell-type proportions}\label{sec::est}
From \eqref{eqn:de}, we estimate the cell-type proportion vector $\bpi_i$ in sample $i$ via solving
the following constrained least-squares problem:
\begin{equation}\label{opt}
    \begin{aligned}
    &\min_{\bpi_i\in\mathbb{R}^{K}} \sum_{j=1}^p\left(y_{ij}-\sum_{k=1}^K\pi_{ik}w_{kj}\right)^2,\\
    &\text{s.t.}\,\,\pi_{ik}\geq 0 \text{  and  } \sum_{k=1}^K\pi_{ik}=1.
    \end{aligned}
\end{equation}
The solution to \eqref{opt} is denoted as $\hat\bpi_i$. Note that $\pi_{ik}\le 1$ is implied by the constrains in \eqref{opt}.
In \eqref{opt}, we consider a constrained ordinary least squares. Alternatively, one may wish to consider a constrained generalized least squares that multiplies the regression equation \eqref{eqn:de} by $\Cov(\bm\epsilon_i)^{-1/2}$. 
While the generalized least squares estimator can be more efficient, we demonstrate in Section \ref{sec:gls} that it may suffer from large biases in practice, due to the uncertainty in estimating $\Cov(\bm\epsilon_i)^{-1/2}$ for each sample $i$.
On the other hand, our empirical investigations show that $\hat\bpi_i$ is more robust and computationally efficient. See detailed discussions and comparisons in Section \ref{sec:gls}.

The optimization problem in \eqref{opt} is a quadratic programming problem and we solve it via the standard dual method \citep{goldfarb1982dual,goldfarb1983numerically}. 
Writing $\W=[\w_1^\top,\ldots,\w_K^\top]\in\mathbb{R}^{p\times K}$, the dual function of \eqref{opt} 
can be written as
\begin{equation}\label{dual}
    \begin{aligned}
    &\max_{\blambda} \c^\top\blambda+\frac{1}{2}(\y_i^\top\y_i-\bpi_i^\top\bW^\top\bW\bpi_i),\\
    &\text{s.t.}\,\,\A^\top
    \blambda+\bW^\top\y_i=(\bW^\top\bW)\bpi_i,
    \end{aligned}
\end{equation}
where $\blambda\in\mathbb{R}^{K+1}$ is the dual vector, $\A=(\I_K,\1_K)^\top\in\mathbb{R}^{(K+1)\times K}$, $\1_K=(1,\ldots,1)\in\mathbb{R}^K$, $\c=(0,\ldots,0,1)^\top\in\mathbb{R}^{K+1}$ and $\bpi_i$ is a solution to \eqref{opt}. Given \eqref{opt} and \eqref{dual}, the standard dual method \citep{goldfarb1982dual,goldfarb1983numerically} that uses the unconstrained least squares estimator as the initial value and the Cholesky
and QR factorizations for parameter updates is applied to calculate $\hat\bpi_i$.

\subsection{Quantifying the uncertainties in estimated proportions}\label{sec::dist}

Write $\Cov(\bm\epsilon_i)=\bSigma_i$ and let $\|\X\|_{1,0}$ denote the number of nonzero entries in a column in $\X$. The next result gives the consistency and asymptotic distribution of $\hat\bpi_i$ from \eqref{opt} and its proof is collected in Section \ref{sec:prooft1}.

\begin{thm}\label{thm2}
Assume that $\bW^\top\bW\in\mathbb{R}^{K\times K}$ is positive definite and as $p\rightarrow\infty$, $\bW^\top\bW/p\rightarrow \bOmega$ and $\bW^\top\bSigma_i\bW/p\rightarrow \G_i$. Suppose $\epsilon_{ij}$'s are sub-exponential random variables, $\|\bSigma_i\|_{1,0}<c$ for some constant $c>0$ and all $\pi_{ik}>0$.
For $1\le i\le n$, it holds that $\hat\bpi_i\overset{P}{\rightarrow}\bpi_i$ 
and 
\begin{equation}\label{asydist}
    \V_i^{-1/2}\sqrt{p}(\hat\bpi_i-\bpi_i)\rightarrow \mathcal{N}(\0,\I),
\end{equation}
where $\V_i=\U\D\U^\top$
and 
$$
\D=\left(\frac{1}{p}\bW^\top\bW\right)^{-1}\left(\frac{1}{p}\bW^\top\bSigma_i\bW\right)\left(\frac{1}{p}\bW^\top\bW\right)^{-1},
$$
$$\U=\I-\left(\frac{1}{p}\bW^\top\bW\right)^{-1}\1_K^\top\left\{\1_K\left(\frac{1}{p}\bW^\top\bW\right)^{-1}\1_K^\top\right\}^{-1}\1_K.
$$
\end{thm}
\noindent

\noindent
Theorem \ref{thm2} shows the asymptotic distribution of the constrained least squares estimator $\hat\bpi_i$. This result is useful for uncertainty quantification in downstream analyses that require cell-type proportions, such as CTS differential expression analysis and CTS eQTL analysis. 
We will demonstrate two such real data examples in Section \ref{sec::rosmap}.
As previously commented, one major challenge in deriving the asymptotic properties of $\hat\bpi_i$ is that 
the random variables $(y_{i1},\ldots, y_{ip})$ in \eqref{eqn:de} are not independent and need to be \textit{decorrelated} using the subject-specific covariance $\bSigma_i$.
In practice, covariance $\bSigma_i$ is unknown and we discuss its estimation in the next section.

Next, in the context of cell type deconvolution analysis, we comment on the plausibility of assumptions made in Theorem \ref{thm2}. The condition that $\bW^\top\bW\in\mathbb{R}^{K\times K}$ is positive definite assumes that the signature gene expressions are not perfectly linearly correlated, which is a mild regularity condition. 
We assume $\epsilon_{ij}$ follows a sub-exponential distribution, which includes the negative binomial distribution that is often used to model read counts from RNA-seq data \citep{robinson2010edger}. The condition on $\|\bSigma_i\|_{1,0}$ assumes that, as the number of signature genes increases, each signature gene only co-expresses with a finite number of other signature genes in each cell type. 
Finally, we assume $\pi_{ik}>0$, that is, all the cell-type proportions in each sample are nonzero, which is reasonable. 
Under this assumption and as $p$ increases, the inequality constraint in \eqref{opt} becomes inactive as $\hat\pi_{ik}>0$ holds with high probability.
When $\pi_{ik}=0$, the distribution of $\hat\pi_{ik}$ in this boundary case becomes very complicated. To simplify, one may instead consider estimating $\bpi_i$ via
\begin{equation*}
\min_{\bpi_i\in\mathbb{R}^{K}} \sum_{j=1}^p\left(y_{ij}-\sum_{k=1}^K\pi_{ik}w_{kj}\right)^2,\,\,\text{s.t.}\,\,\sum_{k=1}^K\pi_{ik}=1,
\end{equation*}
which ignores the nonnegative constraint. The solution to this problem, denoted as $\tilde\bpi_i$, has the same asymptotic distribution as in \eqref{asydist}, and inference made using $\tilde\bpi_i$ and the distribution in \eqref{asydist} is still valid.

\subsection{Estimation of subject-specific covariances}\label{sec::est}

In this section, we discuss the estimation of subject-specific covariance $\bSigma_i$ in \eqref{asydist}. Assuming that the CTS expression profiles $\x_i^{(1)},\ldots,\x_i^{(K)}$ are independent, we may write $\bSigma_i$ as
\begin{equation}\label{ass1eq}
\bSigma_i=\Cov\left(\sum_{k=1}^K\pi_{ik}\x_i^{(k)}\right)=\sum_{k=1}^K\pi_{ik}^2\bSigma^{(k)},
\end{equation}
where $\bSigma^{(k)}=\Cov(\x_i^{(k)})$ represents the CTS co-expression among signature genes. 
In order to estimate $\bSigma_i$, we first focus on the estimation of $\bSigma^{(k)}$.
Centering by $z_{ij}=y_{ij}-\sum_{k=1}^K\pi_{ik}w_{kj}$, it is easy to see that
$$
\mathbb{E}(z_{ij}z_{ij'})=\sum_{k=1}^K \pi_{ik}^2 \bSigma_{jj'}^{(k)},\quad 1\le j,j'\le p.
$$
The above observation facilities an efficient least squares estimation of $(\bSigma_{jj'}^{(1)},\ldots,\bSigma_{jj'}^{(K)})$ by taking $z_{ij} z_{ij'}$ as the response and $(\pi_{i1}^2,\ldots,\pi_{iK}^2)$ as the vector of predictors. 
Writing $\z_{j}=(z_{1j},\ldots,z_{nj})$ and $\H=\left(\pi_{ik}^2\right)_{n\times K}$, the CTS covariances between genes $j$ and $j'$, i.e., $(\bSigma_{jj'}^{(1)},\ldots,\bSigma_{jj'}^{(K)})$, can be consistently estimated with
$$
\b_{jj'}=(\H^\top\H)^{-1}\H^\top(\z_j\circ \z_{j'}),
$$
where $\circ$ denotes the element-wise product. 
The above CTS covariance estimation was first considered by \cite{su2021csnet}, where true cell-type proportions are assumed to be available.

In our setting, the true cell-type proportions are unknown and we only have access to $\hat\H=\left(\hat\pi_{ik}^2\right)_{n\times K}$ and $\hat\z_j$'s, where $\hat z_{ij}=y_{ij}-\sum_{k=1}^K\hat\pi_{ik}w_{kj}$. 
In this case, a natural estimator to consider is  
\begin{equation}\label{sigmakest0}
\hat\b_{jj'}=(\hat\H^\top\hat\H)^{-1}\hat\H^\top(\hat\z_j\circ \hat\z_{j'}).
\end{equation}
Our result in Theorem \ref{thm2} suggests that $\b_{jj'}-\hat\b_{jj'}=O_p(1/\sqrt{p})$.
Hence, $\hat\b_{jj'}$ is also a consistent estimator for $(\bSigma_{jj'}^{(1)},\ldots,\bSigma_{jj'}^{(K)})$ as $p$ increases.

\begin{figure}[!t]
\centering
\includegraphics[trim=0 5mm 0 0, scale=0.6]{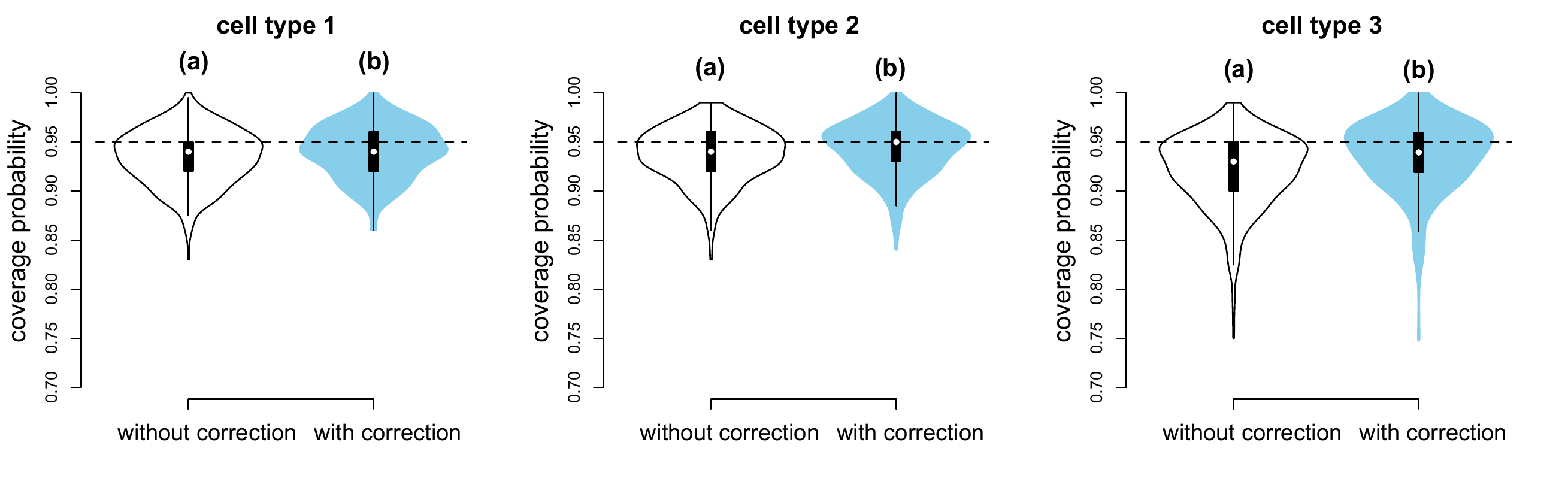}
\caption{\label{fig1} The coverage probabilities of 95\% confidence intervals (CI) in three cell types under the simulation setting in Section \ref{set1}. The (a) \texttt{without correction} CIs are calculated with \eqref{sigmakest0} and the (b) \texttt{with correction} CIs are calculated with \eqref{sigmakest1}.}
\end{figure}

In our empirical studies, we find that the finite-sample bias in $\hat\b_{jj'}$ often leads to a deflated estimation of $\V_i=\Cov(\hat\bpi_i)$ and correspondingly, an under-coverage of the confidence intervals calculated for $\bpi_i$. As an example, Figure \ref{fig1} (a) shows the coverage probabilities of 95\% confidence intervals calculated with \eqref{sigmakest0} under the simulation setting in Section \ref{set1} and some under-coverage is seen.
A further investigation shows that the bias is majorly caused by the finite-sample difference between $\hat\H$ and $\H$ and that between $\hat\H^\top\hat\H$ and $\H^\top\H$. To address this issue, we consider a finite-sample bias-corrected estimator
\begin{equation}\label{sigmakest1}
\hat\b_{jj'}^{\text{correct}}=\left\{\hat{\H}^\top\hat{\H}-\B_1\right\}^{-1}(\hat{\H}-\B_2)^\top(\hat{\z}_j\circ \hat{\z}_{j'}),
\end{equation}
where $\B_1$ and $\B_2$ are calculated by explicitly quantifying $\mathbb{E}(\hat{\H}^\top\hat{\H})-\H^\top\H$ and $\mathbb{E}(\hat{\H})-\H$, respectively, and given in Proposition \ref{lem1} below. The proof is given in Section \ref{sec:proof2}. 

\begin{prop}\label{lem1} 
Letting $\bpi^{\circ 2}_i=(\pi_{i1}^2,\ldots,\pi_{iK}^2)$. If $\sqrt{p}(\hat\bpi_i-\bpi_i)\sim \mathcal{N}(\0,\V_i)$, it holds that
\begin{equation}\label{unbiasedest}
    \begin{split}
    &\bm{B_1}=\frac{1}{p}\sum_{i=1}^n\bpi^{\circ 2}_i \u_i^\top+\frac{1}{p}\sum_{i=1}^n\u_i\bpi^{\circ 2}_i{}^\top+\frac{4}{p}\sum_{i=1}^n(\bpi^{\circ 2}_i{}^\top\bpi^{\circ 2}_i)\circ \V_i+\frac{1}{p^2}\sum_{i=1}^n\T_{i},\\
    &
    \bm{B}_2=\frac{1}{p}[\u_1,\ldots,\u_n]^\top,
    \end{split}
\end{equation}
where $\u_i=(\V_{i,11},\ldots,\V_{i,KK})^\top$ and $\T_{i}$ is a $K\times K$ matrix with 
$T_{i,jj'}=2V_{i,jj'}^2+V_{i,jj}V_{i,j'j'}$.
\end{prop}

\noindent
Based on Proposition \ref{lem1} and given $\V_i$, we can estimate $\B_1$ by
\begin{equation}\label{firstest}
    \hat{\bm{B}}_1=\frac{1}{p}\sum_{i=1}^n{\hat\bpi}^{\circ 2}_i \u_i^\top+\frac{1}{p}\sum_{i=1}^n\u_i{\hat\bpi}^{\circ 2}_i{}^\top+\frac{4}{p}\sum_{i=1}^n({\hat\bpi}^{\circ 2}_i{}^\top{\hat\bpi}^{\circ 2}_i)\circ \V_i+\frac{1}{p^2}\sum_{i=1}^n\T_{i}.
\end{equation} 
As $\V_i$ is unknown in practice, we propose to iteratively update $\bSigma_i$ and $\V_i$ in the estimation procedure. The details are summarized in Algorithm \ref{alg0}. 

\begin{algorithm}
\caption{\label{alg0}The DEcorrelated ConstrAined Least Squares (\texttt{DECALS}) algorithm}
\begin{algorithmic}
\STATE \textbf{Input:} Bulk expressions $\{\bm{y}_i\}_{1\le i\le n}$ and the signature gene matrix $\bm{W}$.\\
\STATE \hspace{0.25in} \textbf{Step 1:} Calculate the constrained least squares estimator $\hat{\bpi}_i$ from \eqref{opt} for $1\le i\le n$.
\STATE \hspace{0.25in} \textbf{Step 2:} Initialize $\V_i^{[0]}$ for $1\le i\le n$.
\STATE \hspace{0.25in} \textbf{Repeat} the following steps for $t=0,1,\ldots$ until convergence.
\STATE \hspace{0.5in} \textbf{Step 3.1:} Calculate $\bm{B}^{[t]}_1$ with \eqref{firstest}, $\hat\bpi_i$ and $\V_i^{[t]}$.
\STATE \hspace{0.5in} \textbf{Step 3.2:} Calculate $(\bSigma^{(k)})^{[t]}$ with \eqref{sigmakest1}, $\hat\H$, $\hat\z_j$ and $\bm{B}^{[t]}_1$.
\STATE \hspace{0.5in} \textbf{Step 3.3:} Calculate $\bSigma_i^{[t]}$ with \eqref{ass1eq}, $\hat\bpi_i$ and $(\bSigma^{(k)})^{[t]}$.
\STATE \hspace{0.5in} \textbf{Step 3.4:} Calculate $\V_i^{[t+1]}$ with \eqref{asydist}, $\W$ and $\bSigma_i^{[t]}$.
\STATE \textbf{Output:} The estimated proportions $\{\hat\bpi_i\}_{1\le i\le n}$ and covariances $\{\hat\V_i\}_{1\le i\le n}$.
\end{algorithmic}
\end{algorithm}
In Step 2 of Algorithm 1, we initialize $\V_i^{[0]}$ by estimating $\sigma^2_i$ with \\
$\sum_{j=1}^p \left(y_{ij}-\sum_{k=1}^K\pi_{ik}w_{kj}\right)^2/(p-1)$. 
When $p$ is large, the accumulated errors across $O(p^2)$ entries in $\hat{\bSigma}^{(k)}$ can be excessive, especially when $p$, the number of signature genes, exceeds $n$, the number of bulk samples. In this case, we consider a sparse estimation of ${\bSigma}^{(k)}$, which is plausible as gene co-expressions are expected to be sparse when $p$ is large \citep{zhang2005general}. Specifically, in Step 3.2 and after calculating $(\bSigma^{(k)})^{[t]}$, we consider a SCAD thresholding procedure \citep{rothman2009generalized} with the tuning parameter selected using cross validation (see Section \ref{sec:sparse}).

\subsection{The constrained generalized least squares}\label{sec:gls}
\begin{figure}[!t]
\centering
\includegraphics[trim=5mm 2cm 0 0,scale=0.615]{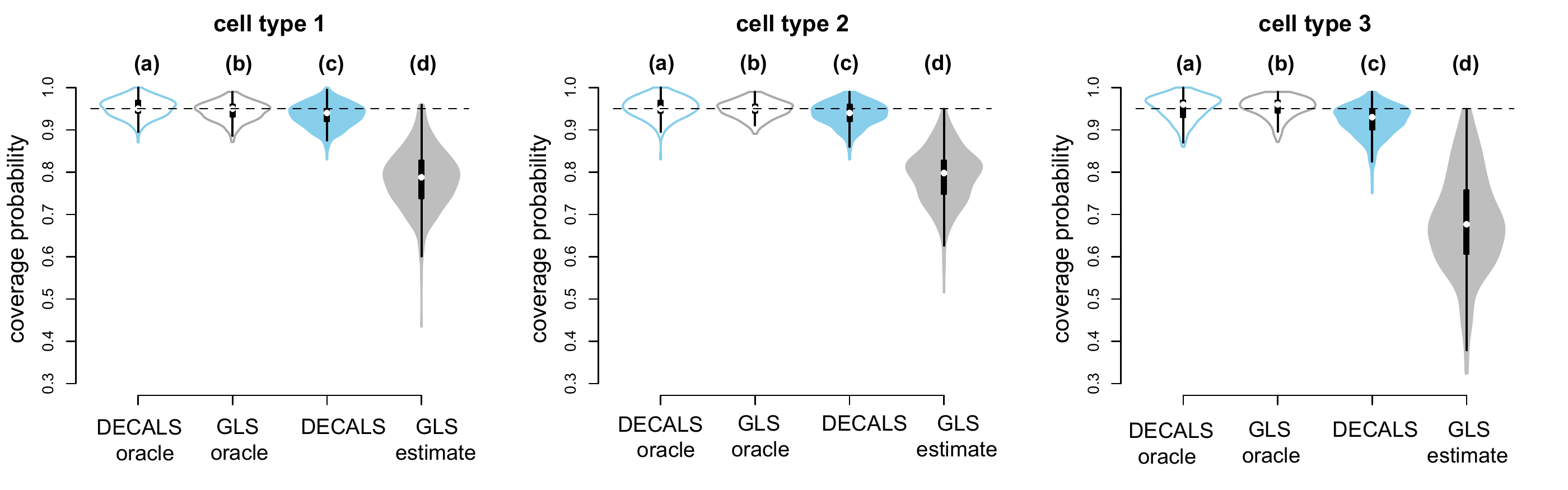}
\caption{\label{fig2} The coverage probabilities of 95\% confidence intervals (CI) for three cell types under the simulation setting in Section \ref{set1}. The (a) \texttt{DECALS oracle} and (c) \texttt{DECALS} CIs are calculated as in Section \ref{sec::dist} with the true and estimated covariance $\V_i$, respectively; the (b) \texttt{GLS oracle} and (d) \texttt{GLS estimate} CIs are calculated with the true and estimated covariance $\V^{\text{GLS}}_i$, respectively.}
\label{fig2}
\end{figure}

In our approach, we estimate $\bpi_i$ via the constrained least squares in \eqref{opt}.
Recalling $\Cov(\bm\epsilon_i)=\bSigma_i$ and assuming $\bSigma_i$ is positive definite, one may prefer to estimate $\bpi_i$ via the following constrained generalized least squares (GLS):
\begin{equation}\label{opt2}
\begin{aligned}
&\min_{\bpi_i\in\mathbb{R}^{k}}\left\Vert\bSigma_i^{-1/2}\y_i-\bSigma_i^{-1/2}\W\bpi_i\right\Vert_2^2,\\
&\text{s.t.}\,\,\pi_{ik}\geq 0 \text{  and  } \sum_{k=1}^K\pi_{ik}=1.
\end{aligned}
\end{equation}
The solution to \eqref{opt2}, denoted as $\hat\bpi_i^{\text{GLS}}$, is expected to be more efficient than $\hat\bpi_i$ \citep{greene2003econometric}. 
Specifically, denoting $\Cov(\hat\bpi_i^{\text{GLS}})$, we have that 
\begin{equation}\label{glsvar}
\V^{\text{GLS}}_i=
(\W^\top\bSigma^{-1}_i\W)^{-1}\left\{\I-\1\{\1^\top(\W^\top\bSigma^{-1}_i\W)^{-1}\1^\top\}^{-1}\1^\top(\W^\top\bSigma^{-1}_i\W)^{-1}\right\}.
\end{equation}
As demonstrated in Section \ref{sec::est}, the estimation of the subject-specific covariance $\bSigma_i$ in our problem is nontrivial. 
When $\bSigma_i$ is unknown but estimated with potentially high noise, the estimate of $\bpi_i^{\text{GLS}}$ from \eqref{opt2} and its variance from \eqref{glsvar}, which further requires $\bSigma_i^{-1}$, can much deteriorate.

As an example, Figure \ref{fig2} shows the coverage probabilities of 95\% confidence intervals calculated with $\hat\bpi_i$ and $\hat\bpi_i^{\text{GLS}}$, respectively, with the true $\bSigma_i$, referred to as {\it oracle}, and estimated $\hat\bSigma_i$ (see details in Section~\ref{gls}) in the simulation setting in Section \ref{set1}.
By comparing plots (a) and (b), it is seen that the GLS estimator is slightly more efficient than \texttt{DECALS} when $\bSigma_i$ is known.
However, when $\bSigma_i$ is unknown and needs to be estimated from data, $\hat\bpi_i^{\text{GLS}}$ and its estimated sampling variance can be biased and the coverage probabilities of the 95\% confidence intervals are unsatisfactory.

\section{Simulation Studies}\label{sec::sim}
We conduct simulations to evaluate the performance of \texttt{DECALS} in two types of settings. In Section \ref{set1}, we generate both the signature gene matrix $\W$ and cell-type proportions $\bpi_i$'s from pre-specified parametric distributions. In Section \ref{set2}, we use the signature gene matrix $\W$ and cell-type proportions $\bpi_i$'s inferred from real data (see more details in Section \ref{set2}). 
Additionally, in Section \ref{sec::noise}, we conduct a sensitivity analysis that examines the performance of \texttt{DECALS} when $\W$ is observed with errors. 

We compare \texttt{DECALS} with three alternative inferential methods including a naive OLS method, referred to as \texttt{OLS}, \texttt{RNA-Sieve} from \citet{erdmann2021likelihood} and \texttt{MEAD} from \citet{xie2022robust}. 
In \texttt{OLS}, the cell-type proportions are estimated from ordinary least squares with no constraints; the proportion estimates are taken to be approximately normal and the covariance is calculated assuming the error terms in \eqref{opt} are i.i.d.. 
\texttt{RNA-Sieve} \citep{erdmann2021likelihood} is a likelihood-based deconvolution method that estimates the cell-type proportions from bulk samples and uses single-cell reference data to infer the distribution of the signature gene matrix. In \texttt{RNA-Sieve}, confidence intervals of the estimated proportions can be calculated as a by-product of the estimation procedure. 
Note that \texttt{RNA-Sieve} requires names of the signature genes to link to single cell reference data and hence, it is only implemented in Section \ref{set2}, where such gene information is available.
\texttt{MEAD} \citep{xie2022robust} uses an error-in-variable regression framework, where the signature gene matrix is estimated from single cell data with noise, to make inference on cell-type proportions. The estimated proportions are shown to be asymptotically normal and the subject-specific covariance is estimated through a sandwich type estimator with an estimated gene-gene dependence set, though this subject-specific covariance estimator may not be consistent. In our implementation of \texttt{MEAD}, we supply the true signature gene matrix without measurement errors and the true gene-gene dependence set under each simulation setting. We compare the performance of these methods through evaluating the coverage probabilities of the confidence intervals constructed by these methods in our experiments.


\subsection{Experiments with simulated \texorpdfstring{$\W$ and $\bpi_i$'s}{TEXT}}
\label{set1}
\begin{figure}[!t]
\centering
\includegraphics[trim=0 5mm 0 0, scale=0.65]{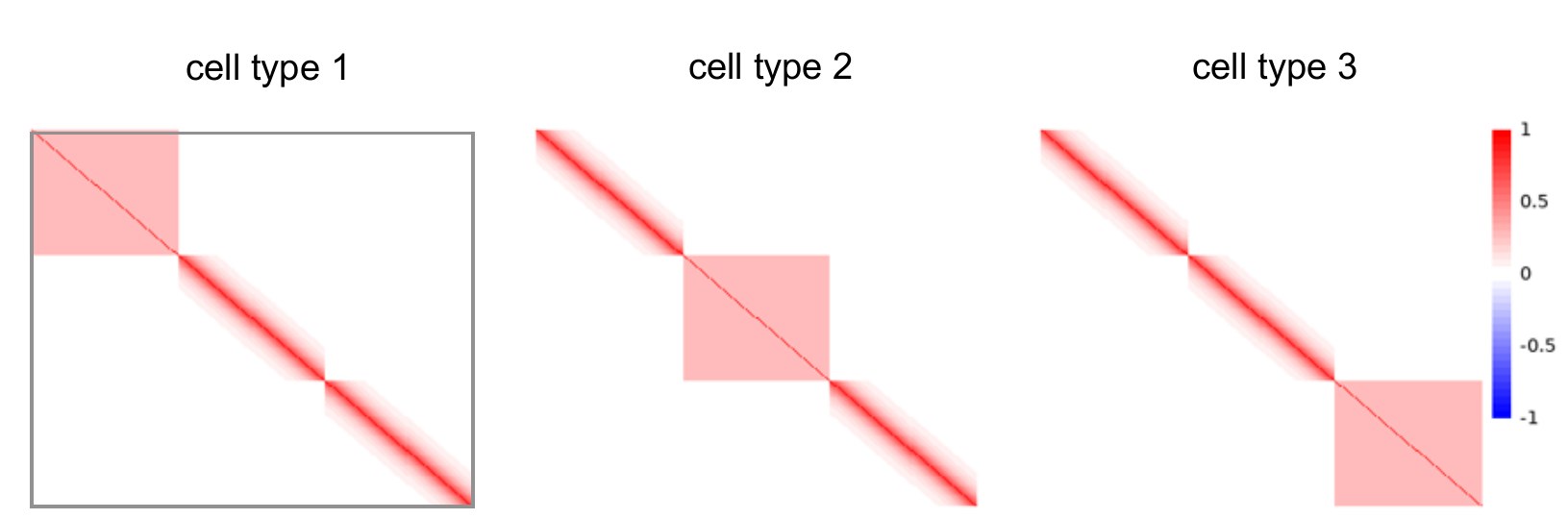}
\caption{An illustration of the CTS correlation matrices in Section \ref{set1}.} 
\label{fig5}
\end{figure}

We consider three cell types $K=3$ and sample $\bpi_i$, the cell-type proportions in sample $i$, from $\bpi_i\sim\text{Dirichlet}(3,2,1)$.
Under this setting, the three cell types have average proportions of 1/2, 1/3 and 1/6, respectively. The bulk gene expression for sample $i$ is calculated as $\y_i=\sum_{k=1}^K\pi_{ik}\x_i^{(k)}$, where the expression profile $\x_i^{(k)}$ is simulated from $\x_i^{(k)}\sim\mathcal{N}(\w_k,\bSigma^{(k)})$. 
Non-Gaussian distributions are considered in Section \ref{set2}.
Entries in $\w_k$ are i.i.d. from $\mathcal{N}(0,1^2)$ and $\bSigma^{(k)}=10\times\R^{(k)}$, where $\R^{(k)}$ is the correlation matrix in cell type $k$.
We let 
$$
\R^{(1)}=\text{diag}(\R_1,\R_2,\R_2);\quad\R^{(2)}=\text{diag}(\R_2,\R_1,\R_2);\quad\R^{(3)}=\text{diag}(\R_2,\R_2,\R_1),
$$
where $\R_1\in\mathbb{R}^{\frac{p}{3}\times \frac{p}{3}}$ with $\R_{1,jj'}=0.3$ and $\R_2\in\mathbb{R}^{\frac{p}{3}\times \frac{p}{3}}$ with $\R_{2,jj'}=0.7\times 0.9^{|j-j'-1|}$, $j\neq j'$; see Figure~\ref{fig5} for an illustration. 
We let the number of signature genes $p=300$ and the number of samples $n=500$.

\begin{figure}[!t]
\centering
\includegraphics[trim=0 20mm 0 0, scale=0.475]{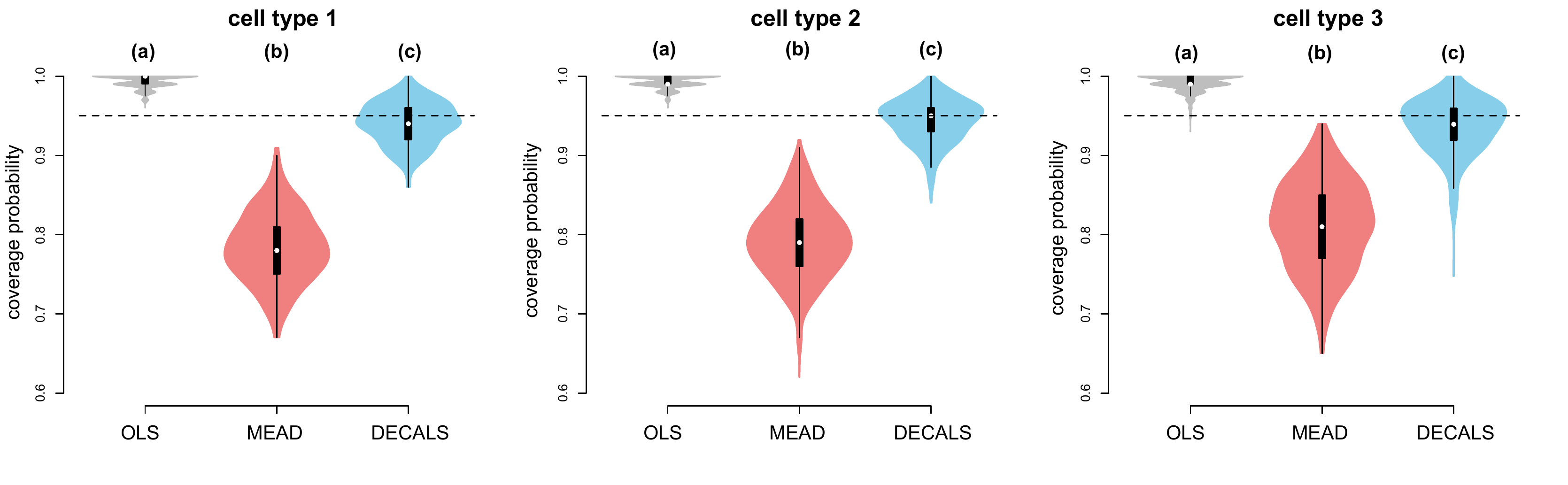}
\caption{\label{fig4} The coverage probabilities of 95\% confidence intervals in three cell types with (a) \texttt{OLS}, (b) \texttt{MEAD} and (c) \texttt{DECALS}.}
\end{figure}

We apply \texttt{OLS}, \texttt{MEAD} and \texttt{DECALS} to infer cell-type proportions for each subject. 
Specifically, we construct 95\% confidence intervals for $\pi_{ik}$'s using each method and estimate the coverage probabilities using 100 data replicates. 
The results are summarized in Figure~\ref{fig4}.
It is seen that \texttt{DECALS} has the best performance for all three cell types, with coverage probabilities close to the nominal level of 95\%. \texttt{OLS} tends to overestimate the CTS proportion variances and the resulting coverage probabilities for the 95\% confidence intervals are consistently greater than 95\%. 
This is majorly because the correlations among signature genes are ignored in \texttt{OLS}. As all CTS correlations are positive in this simulation setting, ignoring these positive correlations inflates the variance estimates of \texttt{OLS}, leading to an over-coverage of the \texttt{OLS} confidence intervals. 
\texttt{MEAD} tends to underestimate the variances for the estimated cell-type proportions, likely because the subject-specific covariance is not consistently estimated using the sandwich estimator. 
Finally, we also investigate the estimation accuracy of $\V_i$ with varying $p$ and $\W$ and the results are shown in Table \ref{tab1} in the supplementary materials. The estimation accuracy of \texttt{DECALS} is satisfactory and it improves with the number of signature genes and the variance of signature gene expressions.


\begin{figure}[!t]
\centering
\includegraphics[trim=0 5mm 0 0, scale=0.75]{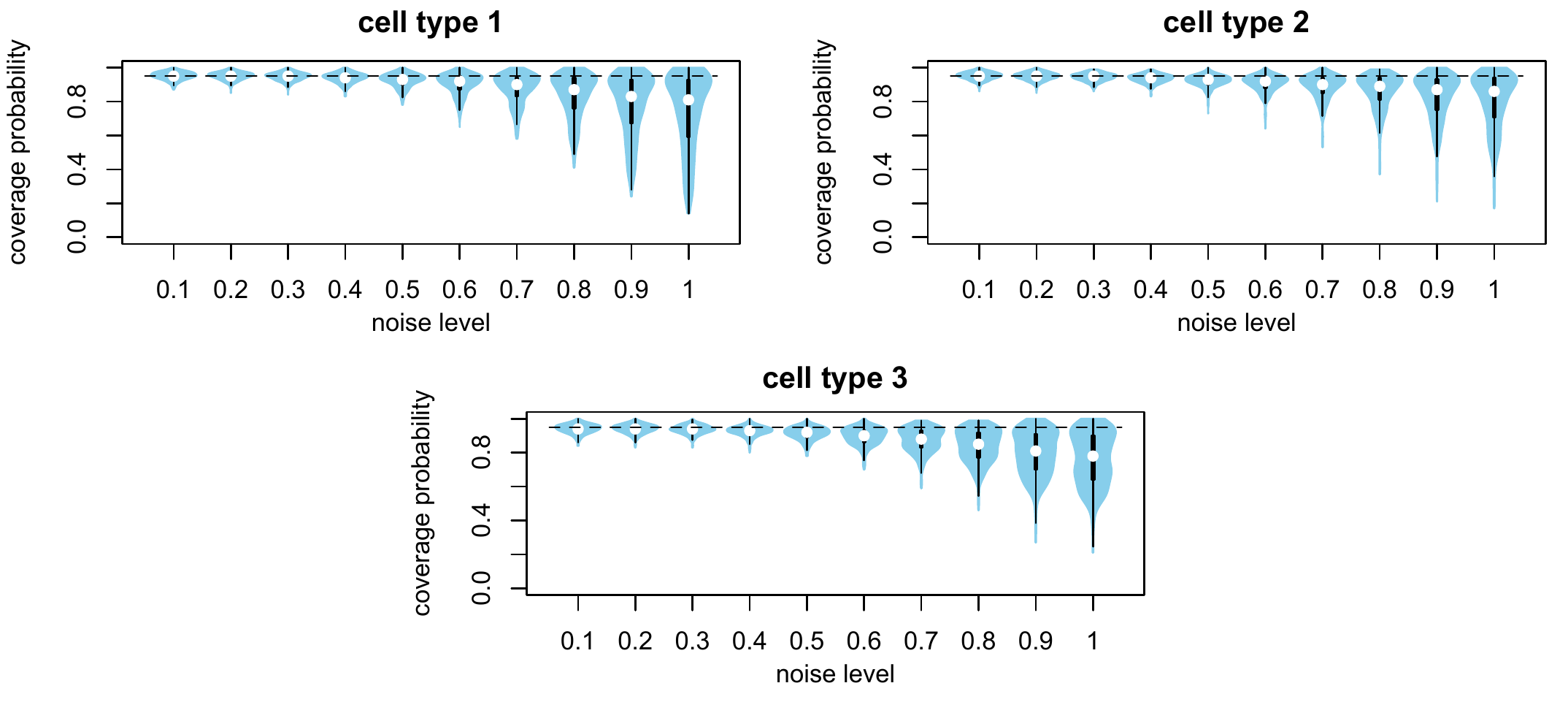}
\caption{\label{noise}The coverage probabilities in three cell types as the noise level $a_0$ varies.}
\end{figure}

\subsection{Sensitivity analysis}\label{sec::noise}
In this section, we conduct a sensitivity analysis to examine the performance of \texttt{DECALS} when the signature gene expression matrix $\W$ is inaccurate and observed with errors.
Consider the simulation settings in Section \ref{set1}, where 
the mean signature gene expression is generated using $w_{kj}\overset{i.i.d.}{\sim} N(0,1^2)$. In this sensitivity analysis, we assume that instead of $w_{kj}$, we observe $\tilde{w}_{kj}=w_{kj}+e_{kj}$, where $e_{kj}\overset{i.i.d.}{\sim} N(0,a_0^2)$ and $a_0$ is set to be between 0.1 and 1 with a step size 0.1. 
Figure \ref{noise} reports the coverage probabilities of 95\% confidence intervals with \texttt{DECALS} under various noise levels. It is seen that under this inaccurate signature gene matrix setting, \texttt{DECALS} still performs reasonably well, with the coverage probabilities close to the nominal level of 95\% when $a_0$ is as large as 0.6.

\subsection{Experiments with \texorpdfstring{$\W$ and $\bpi_i$'s}{TEXT} inferred from real data}\label{set2}
For experiments in this section, we use the signature gene matrix $\W$, cell-type proportions $\bpi_i$'s and CTS covariances $\bSigma^{(k)}$'s inferred from the real data analysis in Section \ref{sec::rosmap}. 
There are $K=5$ five cell types in this dataset, $n=541$ bulk samples and $p=159$ signature genes. 

\begin{figure}[!t]
\centering
\includegraphics[trim=0 10mm 0 0, scale=0.75]{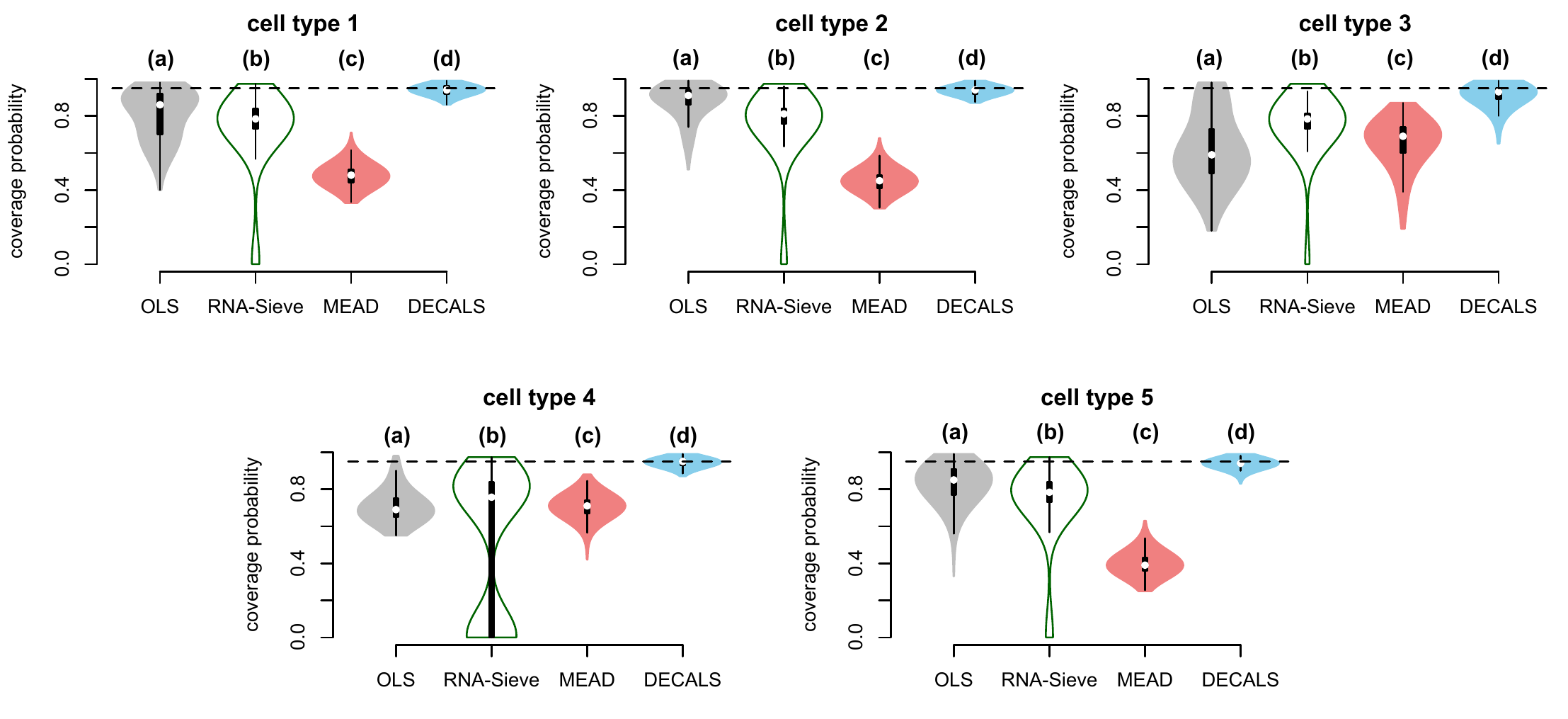}
\caption{The coverage probabilities of 95\% confidence intervals with (a) \texttt{OLS}, (b) \texttt{RNA-Sieve}, (c) \texttt{MEAD} and (d) \texttt{DECALS}.}
\label{result3}
\end{figure}

As the dataset in Section \ref{sec::rosmap} uses expression unit FPKM \citep{trapnell2010transcript} to measure gene expression, which is continuous and positive, we generate CTS expression profile $x_{ij}^{(k)}$'s from Gamma distributions. 
Specifically, given the mean $\w_k$ and target covariance $\bSigma^{(k)}$ inferred from real data, we simulate $\x_i^{(k)}$ using a copula approach (see Section \ref{sec::gamma} in the supplement), similar to that in \citet{tian2021esco}.
We apply \texttt{OLS}, \texttt{MEAD}, \texttt{RNA-Sieve} and \texttt{DECALS} to infer cell-type proportions for each subject. Specifically, we construct 95\% confidence intervals for $\pi_{ik}$'s by each method and estimate the coverage probabilities using 100 data replicates. 
The results are summarized in Figure \ref{result3}.
It is seen that \texttt{DECALS} has the best performance in all five cell types, with coverage probabilities close to the nominal level of 95\%. 
Similar as before, \texttt{MEAD} tends to underestimate the variance for the estimated cell-type proportions, which results in confidence intervals with under-coverage.
It is seen that \texttt{RNA-Sieve} also suffers from under-coverage, which could be due to violations of key \texttt{RNA-Sieve} model assumptions. For example, \texttt{RNA-Sieve} assumes that the expression levels in different genes are independent and Gaussian distributed, which does not hold under this simulation setting. We also simulate gene expression data from Gaussian distributions. The results are similar and can be found in Section \ref{normal} in the supplement.


\section{Using DECALS in CTS Analysis from Bulk Samples}\label{sec::real}
\subsection{A sampling approach to incorporating uncertainties
}\label{sec::boot}


As mentioned above, most CTS analysis methods using bulk samples require cell-type proportions across samples as input, including methods that infer CTS gene expressions \citep{wang2021bayesian}, CTS differentially expressed genes \citep{jin2021cell,wang2021bayesian}, CTS eQTLs \citep{patel2021cell,little2022cell} and CTS co-expressions \citep{su2021csnet}. As the cell-type proportions used in these methods are not known but estimated from bulk sample data, incorporating the uncertainties in the estimated proportions with \texttt{DECALS} can mitigate the potential bias resulting from treating the proportions as known, a common assumption made in the existing CTS methods, and lead to more accurate and biologically more interpretable findings.

One possible approach to incorporating the inferred uncertainties for a specific CTS analysis method is to repeatedly sample the cell-type proportions from the distributions of $\hat\bpi_i$'s inferred from \texttt{DECALS} and perform analysis for each set of these sampled proportions. We can then summarize the results across these repeats. More specifically, we sample $M$ sets of proportions denoted as $\{\hat\bpi_i^{[m]}\}_{1\le i\le n}$ for $m=1,\ldots,M$. For each set of sampled proportions $\{\hat\bpi_i^{[m]}\}_{1\le i\le n}$, we apply the CTS analysis method and get an output, denoted as $\mathcal{S}^{[m]}$. Here $\mathcal{S}^{[m]}$ can be CTS gene expression estimates or a set of CTS differentially expresssed genes. With the results $\mathcal{S}^{[1]},\ldots,\mathcal{S}^{[M]}$ from $M$ sets of sampled proportions, we can incorporate uncertainty from cell-type proportion estimates in the CTS analysis method via, for example, computing confidence intervals.

In Sections \ref{sec::rosmap} and \ref{sec::gtex}, we implement the above procedure by applying \texttt{DECALS} to two real data sets to infer uncertainties associated with cell-type proportion estimates, and incorporate these uncertainties in downstream analysis that identifies differentially expressed genes/transcripts in a specific cell type. 
Specifically, we combine \texttt{DECALS} with a downstream CTS analysis method in \citet{wang2021bayesian}, referred to as \texttt{bMIND}, that uses bulk sample data and cell-type proportions to identify CTS differentially expressed genes/transcripts. 
\texttt{bMIND} adopts a Bayesian approach to estimating CTS expressions from bulk RNA-seq data, which are then used to detect CTS differentially expressed genes/transcripts. The method takes the bulk RNA-seq data and cell-type proportions across samples as the input and outputs the set of genes/transcripts inferred to be differentially expressed in each cell type. 
We show that, by considering uncertainties of the estimated proportions in \texttt{bMIND}, we can get results that are more enriched for biologically relevant functions and more interpretable.

\subsection{ROSMAP Data}\label{sec::rosmap}
\begin{figure}[!t]
\centering
\includegraphics[trim=0 20mm 0 0 ,scale=0.75]{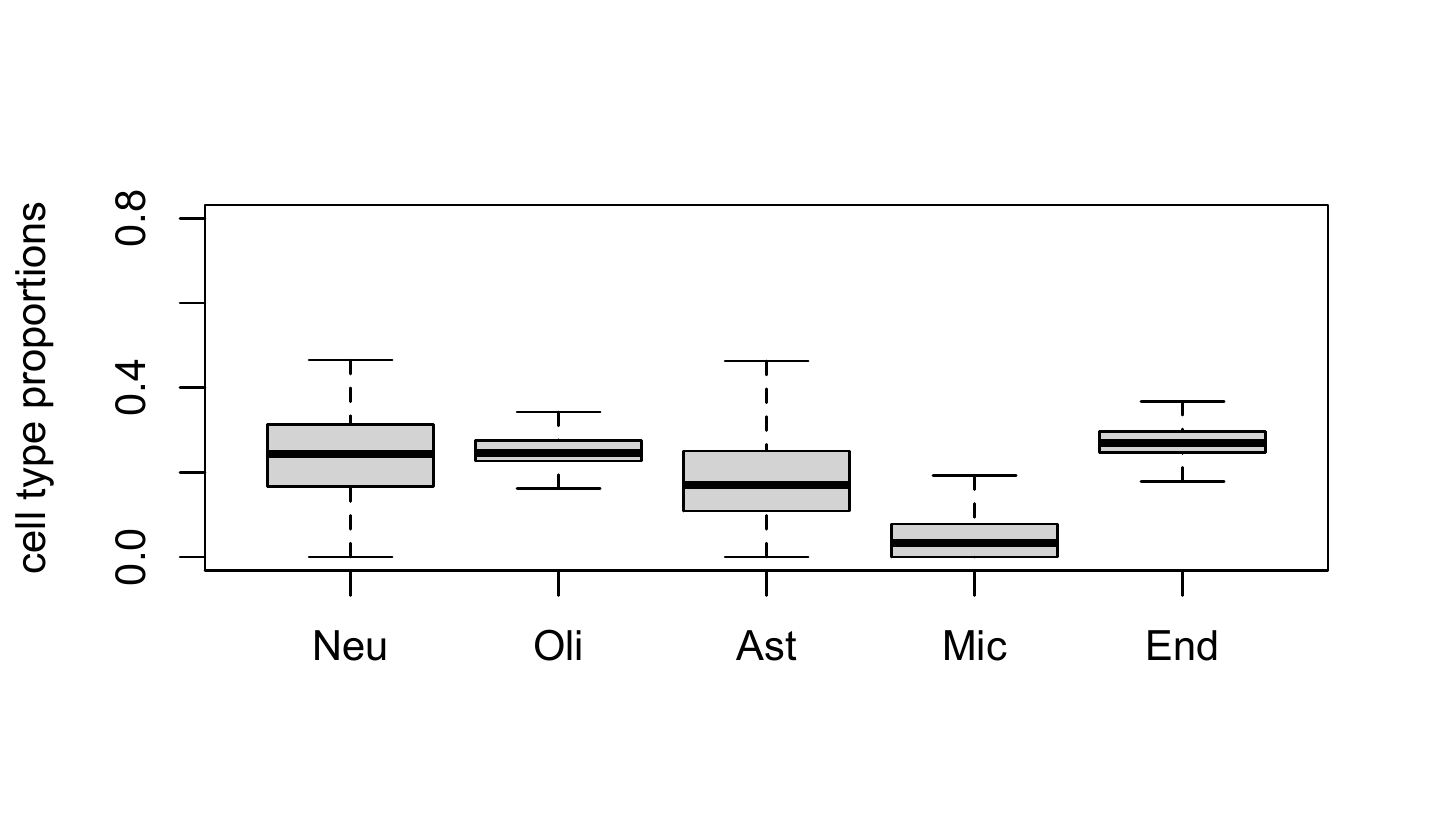}
\caption{Estimated cell-type proportions across all samples in the ROSMAP data.}
\label{CTSp}
\end{figure}

We consider the bulk RNA-seq data collected from the Religious Orders Study and Rush Memory and Aging Project \citep[ROSMAP;][]{bennett2018religious}, a clinical-pathologic cohort study of Alzheimer’s disease. Post-mortem brain samples from $n=541$ subjects were collected from the dorsolateral prefrontal cortex, a brain region that is strongly associated with Alzheimer’s disease pathology \citep{salat2001selective,montembeault2016altered}. 
Among the 541 subjects, 219 were Alzheimer's disease patients and 322 were controls. 
The bulk gene expression levels for each subject were collected in \citet{mostafavi2018molecular}\footnote{\url{https://www.synapse.org/\#!Synapse:syn3388564}} and measured in units of FPKM \citep{trapnell2010transcript}. 
Using the single-nucleus RNA-seq data from \cite{mathys2019single}\footnote{\url{https://www.synapse.org/\#!Synapse:syn21261143}}, we applied the CIBERSORTx S-mode \citep{newman2019determining} to correct for batch effects and 
to obtain a candidate signature gene matrix for five major cell types,
including neurons (Neu), oligodendrocytes (Oli), astrocytes (Ast), microglia (Mic), and endothelials (End). 
To ensure that the final selected signature genes had strong differential signals across these five cell types, we further took the intersection of this candidate gene set with the differentially expressed marker genes for each cell type from \cite{mathys2019single}, which finally gave a signature matrix with $p=159$ genes for the five cell types.
Given the bulk sample expressions and the signature gene matrix, we applied \texttt{DECALS} to estimate the cell-type proportions and their sampling distributions. Figure~\ref{CTSp} presents the estimated proportions across 541 samples, which shows a good agreement with the cell-type abundances reported in \cite{patrick2020deconvolving} on a subset of ROSMAP samples. 

Next, we focused a set of 9,328 protein coding genes with FPKM$\ge4$ in at least half of the samples, and applied \texttt{bMIND} to identify CTS differentially expressed (DE) genes between Alzheimer's disease patients and controls. 
We considered two different approaches to inferring DE genes. 
The first approach directly applied \texttt{bMIND} with $\hat\bpi_i$'s estimated from \eqref{opt} and calculated a $p$-value for each gene $j$ in each cell type $k$, denoted as $p_{jk}$.
In \texttt{bMIND}, gene $j$ is considered a DE gene in cell type $k$ if $p_{jk}<0.05$.
The second approach combines \texttt{bMIND} with \texttt{DECALS} as described in Section \ref{sec::boot}. More specifically, given the estimated sampling distributions of $\hat\bpi_i$'s from \texttt{DECALS}, we sampled 100 sets of proportions denoted as $\{\hat\bpi_i^{[m]}\}_{1\le i\le n}$ for $m=1,\ldots,100$. For each set of sampled proportions $\{\hat\bpi_i^{[m]}\}_{1\le i\le n}$, we applied \texttt{bMIND} and calculated the $p$-value for gene $j$ in cell type $k$, denoted as $p_{jk}^{[m]}$. 
After 100 repeats, gene $j$ was considered a DE gene in cell type $k$ if $\sum_{m=1}^{100}1\{p_{jk}^{(m)}<0.05\}>10$, where the cut-off value of 10 was calculated as two standard deviations above the expected value of $\sum_{m=1}^{100}1\{p_{jk}^{(m)}<0.05\}$ for a non DE gene. Specifically, the number of times a non DE gene is selected follows a Binomial(100,0.05), with a mean 5 and variance 4.75.
We refer to these two approaches as \texttt{bMIND} and \texttt{bMIND+DECALS}, respectively.

To compare the DE gene sets identified from the two approaches for each of the five cell types, we performed enrichment analysis using QIAGEN Ingenuity Pathway Analysis (IPA, QIAGEN Inc., \url{https://digitalinsights.qiagen.com/IPA}). IPA identifies pathways enriched with DE genes by testing the association between the input genes and canonical pathways by first calculating the ratio of the number of genes in the input gene set that map to each pathway, and then using a Fisher’s exact test to assess the statistical significance for the association between the input gene sets and canonical pathways \citep{ipa}. We hypothesized that as \texttt{bMIND+DECALS} considered uncertainties in the cell-type proportion estimates, the inferred DE gene sets should be more enriched with biological signals as reflected from the IPA analysis. Because a larger gene set is likely more enriched for biological signals, when the gene sets inferred from \texttt{bMIND} and \texttt{bMIND+DECALS} differed in size, we only kept the top significant genes from the method with the larger gene set so that the resulting two gene sets had the same size in the enrichment analysis.

\begin{table}[!t]
\centering
\begin{tabular}{c|ccccc}\hline
& Neu & Oli & Ast & Mic & End  \\\hline
\texttt{bMIND} & 0 & 0 & 6 & 19 & 0  \\
\texttt{bMIND+DECALS} & 1 & 7 & 5 & 55 & 0 \\\hline
\end{tabular}
\caption{Numbers of pathways selected in the IPA enrichment analysis.}
\label{ipa}
\end{table}

\textbf{Enriched biological findings from \texttt{bMIND+DECALS}}.
Table~\ref{ipa} shows that the \texttt{bMIND+DECALS} approach implied a larger number of significant IPA canonical pathways than the \texttt{bMIND} approach in most cell types (see significant IPA canonical pathways in Section \ref{sec::pathways} in the supplementary materials).
This result suggests that the DE gene sets identified from the \texttt{bMIND+DECALS} procedure can potentially offer more biological insights than those from the \texttt{bMIND} procedure. Moreover, a further investigation shows that \texttt{bMIND+DECALS} might better identify canonical pathways related to Alzheimer's disease. 
For instance, in oligodendrocyte (Oli), 
the Sumoylation pathway was only identified in \texttt{bMIND+DECALS} (Benjamini–Hochberg adjusted [BH] p-value = $3.24\times 10^{-7}$). This pathway was previously reported to regulate amyloid precursor proteins, which are central to Alzheimer’s disease \citep{li2003positive,martin2007emerging,anderson2017sumoylation}. 
In astrocytes (Ast), the top three pathways identified in \texttt{bMIND+DECALS} were EIF2 Signaling (BH p-value = $1.16\times 10^{-25}$), mTOR Signaling (BH p-value = $1.07\times 10^{-8}$) and Regulation of eIF4 and p70S6K Signaling (BH p-value = $2.40\times 10^{-8}$). These three pathways were previously reported to have associations with the development of Alzheimer's disease through meta-analysis \citep{adpathways}. Specifically, mTOR Signaling, which was already shown to be associated with Alzheimer's disease \citep{congdon2018tau,butterfield2019oxidative}, was not identified by \texttt{bMIND}. 
In microglia (Mic), Cholesterol Biosynthesis I (BH p-value = $2.63\times 10^{-5}$), Cholesterol Biosynthesis II (BH p-value = $2.63\times 10^{-5}$), Cholesterol Biosynthesis III (BH p-value = $2.63\times 10^{-5}$) and Putrescine Degradation III (BH p-value = $7.08\times 10^{-5}$) pathways were only identified by \texttt{bMIND+DECALS}. These pathways were reported to be related to amyloid-$\beta$ peptide \citep{reitz2011epidemiology,chun2020severe}, which is important in Alzheimer's disease. These results further demonstrate the benefit of considering uncertainties in cell-type proportion estimates in identifying CTS DE genes.


\subsection{GTEx Data}\label{sec::gtex}
\begin{figure}[!t]
\centering
\includegraphics[trim=0 20mm 0 0 ,scale=0.75]{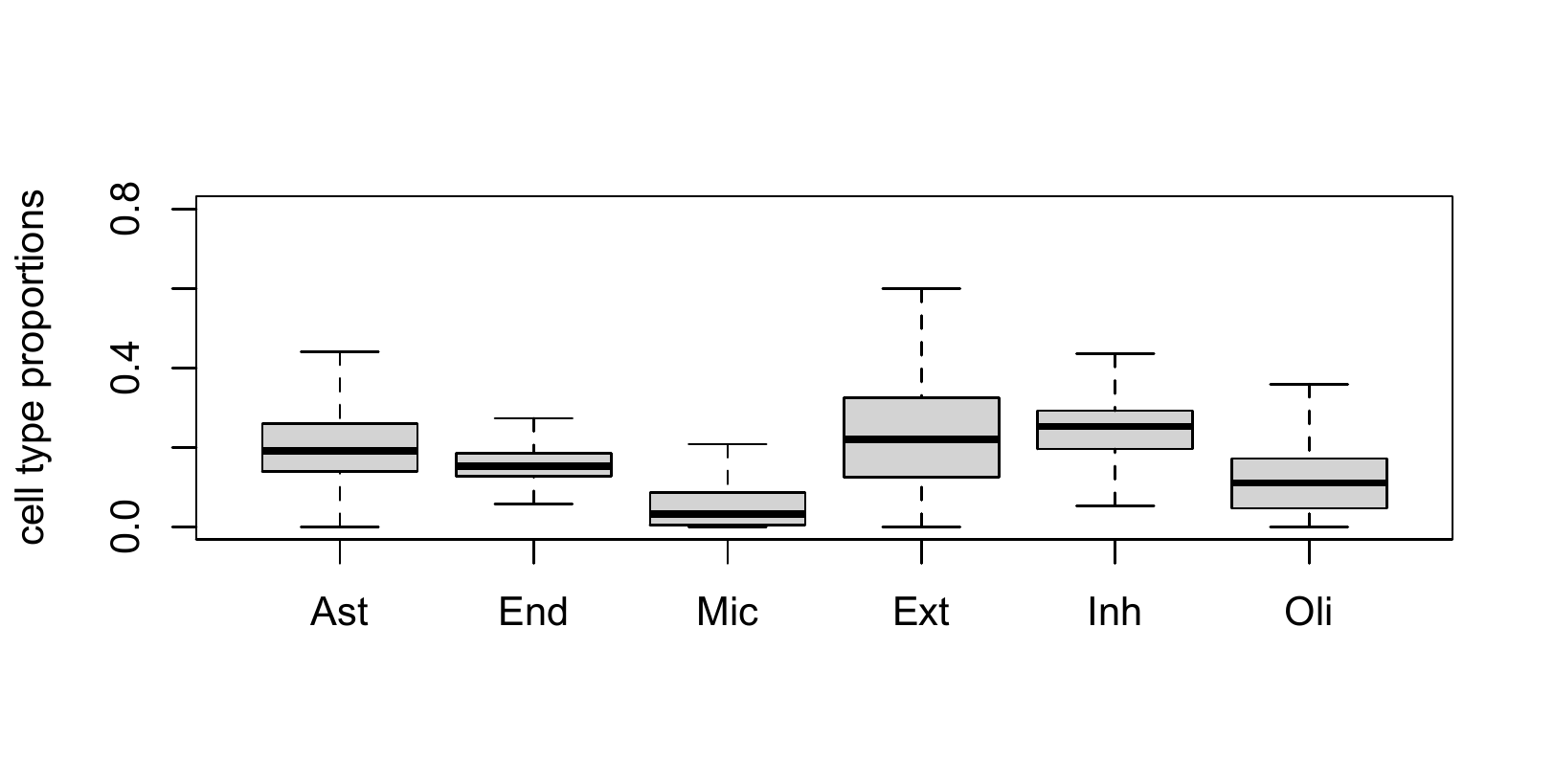}
\caption{Estimated cell-type proportions across all samples in the GTEx data.}
\label{CTSp1}
\end{figure}

The Genotype-Tissue Expression (GTEx) project \citep{gtex2020gtex} is a major effort to collect gene expression data from post-mortem donor samples at a number of non-diseased tissue sites. Our analysis focused on identifying CTS DE transcripts between males and female in brain tissues, as considered in \cite{wang2021bayesian}. 
We focused on $n=1671$ brain samples from the GTEx project (\url{https://www.ncbi.nlm.nih.gov/gap/}) and considered six cell types, including astrocyte (Ast), endothelial (End), microglia (Mic), excitatory (Ext) neuron, inhibitory (Inh) neuron and oligodendrocyte (Oli), and the same set of $p=754$ signature genes as in \cite{wang2021bayesian}. The estimated cell-type proportions from \texttt{DECALS} are shown in Figure~\ref{CTSp1}, consistent with those reported in \cite{wang2021bayesian}.


\begin{figure}[!t]
\centering
\includegraphics[trim=0mm 20mm 0mm 0mm ,scale=0.65]{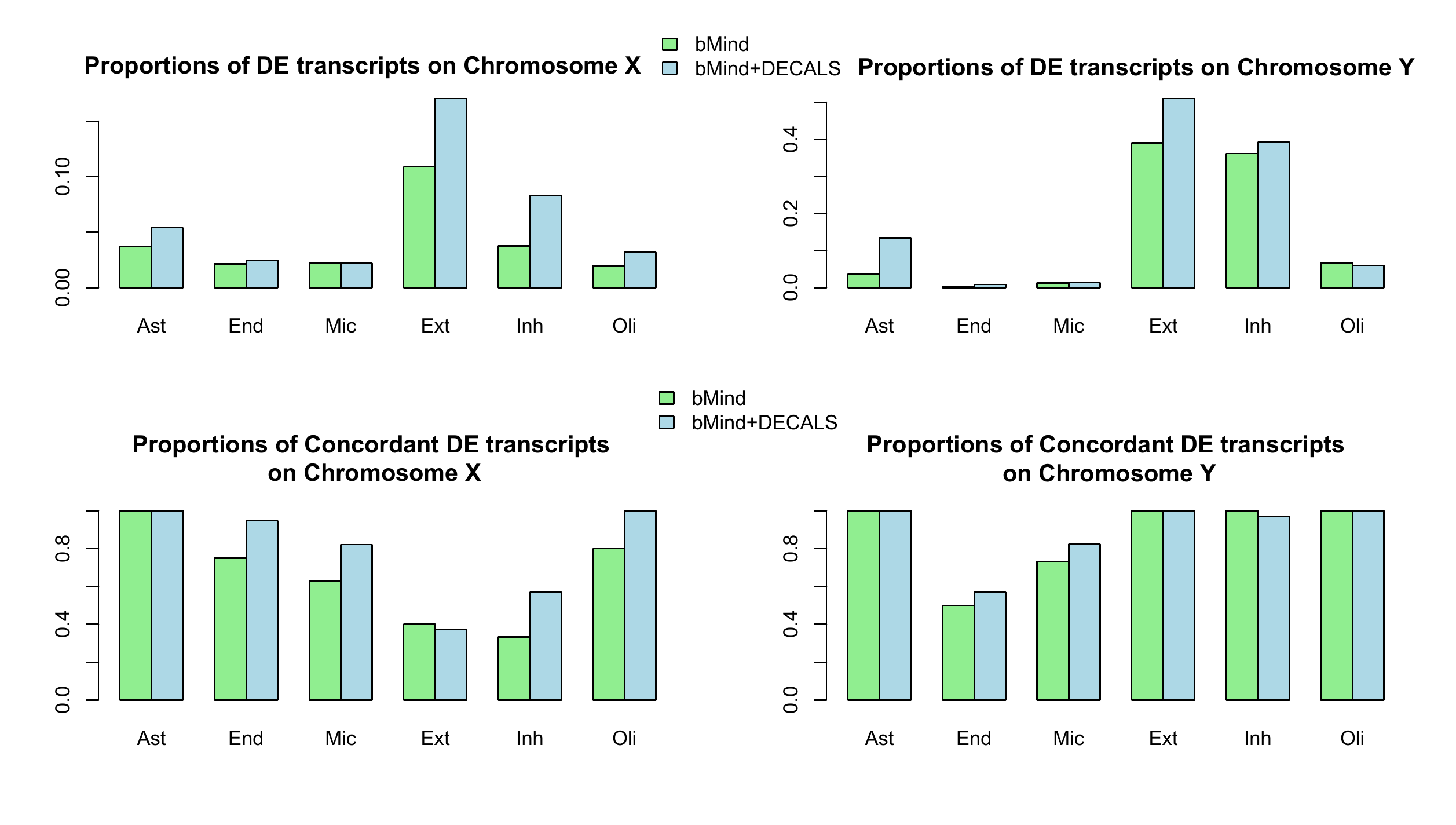}
\caption{\label{GTExratio} Identified CTS DE transcripts using GTEx data 
with proportions of CTS DE transcripts on Chromosomes X and Y (top panel) and proportions of concordant CTS DE transcripts on Chromosomes X and Y, respectively (bottom panel).} 
\end{figure}


\textbf{More biologically interpretable findings from \texttt{bMIND+DECALS}}.
There are a total of 54,271 transcripts in the GTEx dataset and we consider all of them in our analysis. 
Following the procedure in Section \ref{sec::rosmap}, we identified CTS DE transcript sets using \texttt{bMIND} and \texttt{bMIND+DECALS}, respectively. 
We mapped these DE transcripts to all chromosomes including sex chromosomes X and Y. For a method that can better detect DE transcripts between males and females, we would expect larger proportions of the identified DE transcripts on the sex chromosomes. For a transcript set $\mathcal{A}$, denote $\mathcal{A}^X$ and $\mathcal{A}^Y$ as the subsets of $\mathcal{A}$ that are mapped to chromosomes X and Y, respectively. We calculate the proportions of DE transcripts that are mapped to the sex chromosomes as $|\mathcal{A}^X|/|\mathcal{A}|$ and $|\mathcal{A}^Y|/|\mathcal{A}|$, where $|\cdot|$ denotes the cardinality of a set. The top panel of Figure \ref{GTExratio} shows that \texttt{bMIND+DECALS} has higher proportions of DE transcripts mapped to the sex chromosomes in most cell types than \texttt{bMIND}.

Next, we compare the concordance of DE transcripts on sex chromosomes.
Specifically, it is expected that females will more likely have higher expression levels for DE transcripts on the X chromosome and males will more likely have higher expression levels for DE transcripts on the Y chromosome. Correspondingly, if a DE transcript on the X (Y) chromosome is over-expressed in females (males), we referred to it as a \textit{concordant} DE transcript.
For the set of DE transcripts $\mathcal{A}^X$ mapped to the chromosome X, we denote $\mathcal{A}^{XF}$ and $\mathcal{A}^{XM}$ as the subsets of $\mathcal{A}^X$ that are over-expressed in females and males, respectively. Similarly, for the set of DE transcripts $\mathcal{A}^Y$ mapped to the Y chromosome, we can define subsets $\mathcal{A}^{YF}$ and $\mathcal{A}^{YM}$. 
We calculate the proportions of concordant DE transcripts as $|\mathcal{A}^{XF}|/|\mathcal{A}^X|$ and $|\mathcal{A}^{YM}|/|\mathcal{A}^Y|$, respectively.
The bottom panel of Figure \ref{GTExratio} shows that the majority of CTS DE transcripts identified by both methods are concordant. Moreover, \texttt{bMIND+DECALS} has higher proportions of concordant DE transcripts in most cell types, further suggesting considering the uncertainties in the estimated cell-type proportions can improve the detection of CTS DE transcripts.

\section{Discussion}\label{sec:discsuss}
We have proposed a decorrelated constrained least squares (\texttt{DECALS}) framework that estimates cell-type proportions as well as their sampling distributions under a flexible statistical deconvolution framework that allows a general and subject-specific covariance of bulk gene expressions. 
We demonstrate through the analyses of bulk gene expression data from post mortem brain samples that considering the uncertainties in the estimated cell type proportions can lead to more enriched and interpretable biological findings in downstream CTS analysis. 
Our proposed method \texttt{DECALS} is flexible, easy to compute and can be combined with most CTS analysis methods using bulk samples, such as CTS gene expression and co-expression estimation, CTS DE gene and eQTL identification, to improve the accuracy and interpretability of the results.

Our approach assumes that the mean expression levels of signature genes in $\W$ are given. 
In cell type deconvolution analysis, 
$\W$ is usually gathered from pure cell types \citep{newman2015robust, li2016comprehensive} or single cell RNA-sequencing data \citep{wang2019bulk, newman2019determining, jew2020accurate}. 
Our empirical investigations showed that \texttt{DECALS} is not sensitive to errors in $\W$ (see Section \ref{sec::noise}). As a future direction, it is possible to further extend our framework to accommodate a noisy $\W$ by formulating \eqref{eqn:de} as a measurement error model, similar to that in \citet{xie2022robust}. 
The errors in $\W$ can possibly be quantified via modeling the scRNA-seq data. We leave the full investigation of this topic as future research.

\section*{Acknowledgements}
We thank the ROSMAP team for their permission, requested at \url{https://www.radc.rush.edu}, to access the bulk RNA-seq and single nueclues RNA-seq data in the project. The ROSMAP project is supported by the following grants: P30AG72975, P30AG010161 (ADCC), R01AG015819 (RISK), R01AG017917 (MAP), U01AG46152 (AMP-AD Pipeline I) and\\ U01AG61356 (AMP-AD Pipeline II). Zhang was supported by NSF grant DMS 2210469. Zhao was supported by NIH grants R01 GM134005 and R56 AG074015.

\bibliographystyle{asa}
\begingroup
\baselineskip=17.5pt
\bibliography{main.bbl}

\begin{thebibliography}{56}
\newcommand{\enquote}[1]{``#1''}
\expandafter\ifx\csname natexlab\endcsname\relax\def\natexlab#1{#1}\fi

\bibitem[{Abbas et~al.(2009)Abbas, Wolslegel, Seshasayee, Modrusan, and
  Clark}]{abbas2009deconvolution}
Abbas, A.~R., Wolslegel, K., Seshasayee, D., Modrusan, Z., and Clark, H.~F.
  (2009), \enquote{Deconvolution of blood microarray data identifies cellular
  activation patterns in systemic lupus erythematosus,} \textit{PloS one}, 4,
  e6098.

\bibitem[{Anderson et~al.(2017)Anderson, Zanella, Henley, and
  Cimarosti}]{anderson2017sumoylation}
Anderson, D.~B., Zanella, C.~A., Henley, J.~M., and Cimarosti, H. (2017),
  \enquote{Sumoylation: implications for neurodegenerative diseases,}
  \textit{SUMO Regulation of Cellular Processes}, 261--281.

\bibitem[{Barab{\'a}si et~al.(2011)Barab{\'a}si, Gulbahce, and
  Loscalzo}]{barabasi2011network}
Barab{\'a}si, A.-L., Gulbahce, N., and Loscalzo, J. (2011), \enquote{Network
  medicine: a network-based approach to human disease,} \textit{Nature reviews
  genetics}, 12, 56--68.

\bibitem[{Bennett et~al.(2018)Bennett, Buchman, Boyle, Barnes, Wilson, and
  Schneider}]{bennett2018religious}
Bennett, D.~A., Buchman, A.~S., Boyle, P.~A., Barnes, L.~L., Wilson, R.~S., and
  Schneider, J.~A. (2018), \enquote{Religious orders study and rush memory and
  aging project,} \textit{Journal of Alzheimer's disease}, 64, S161--S189.

\bibitem[{Bradley et~al.(2015)Bradley, Holan, and
  Wikle}]{bradley2015multivariate}
Bradley, J.~R., Holan, S.~H., and Wikle, C.~K. (2015), \enquote{Multivariate
  spatio-temporal models for high-dimensional areal data with application to
  longitudinal employer-household dynamics,} \textit{The Annals of Applied
  Statistics}, 9, 1761--1791.

\bibitem[{Butterfield and Halliwell(2019)}]{butterfield2019oxidative}
Butterfield, D.~A. and Halliwell, B. (2019), \enquote{Oxidative stress,
  dysfunctional glucose metabolism and Alzheimer disease,} \textit{Nature
  Reviews Neuroscience}, 20, 148--160.

\bibitem[{Chen and Shao(2004)}]{chen2004normal}
Chen, L.~H. and Shao, Q.-M. (2004), \enquote{Normal approximation under local
  dependence,} \textit{The Annals of Probability}, 32, 1985--2028.

\bibitem[{Chen et~al.(2009)Chen, Garcia, Gupta, Rahimi, and
  Cazzanti}]{chen2009similarity}
Chen, Y., Garcia, E.~K., Gupta, M.~R., Rahimi, A., and Cazzanti, L. (2009),
  \enquote{Similarity-based classification: Concepts and algorithms.}
  \textit{Journal of Machine Learning Research}, 10.

\bibitem[{Chun et~al.(2020)Chun, Im, Kang, Kim, Shin, Won, Lim, Ju, Park, Kim,
  et~al.}]{chun2020severe}
Chun, H., Im, H., Kang, Y.~J., Kim, Y., Shin, J.~H., Won, W., Lim, J., Ju, Y.,
  Park, Y.~M., Kim, S., et~al. (2020), \enquote{Severe reactive astrocytes
  precipitate pathological hallmarks of Alzheimer’s disease via H2O2-
  production,} \textit{Nature neuroscience}, 23, 1555--1566.

\bibitem[{Congdon and Sigurdsson(2018)}]{congdon2018tau}
Congdon, E.~E. and Sigurdsson, E.~M. (2018), \enquote{Tau-targeting therapies
  for Alzheimer disease,} \textit{Nature Reviews Neurology}, 14, 399--415.

\bibitem[{Consortium(2020)}]{gtex2020gtex}
Consortium, G. (2020), \enquote{The GTEx Consortium atlas of genetic regulatory
  effects across human tissues,} \textit{Science}, 369, 1318--1330.

\bibitem[{Dantzig and Cottle(1968)}]{dantzig1968complementary}
Dantzig, G. and Cottle, R. (1968), \enquote{Complementary pivot theory of.
  mathematical programming,} \textit{Mathematics of the decision sciences,
  part}, 1, 115--136.

\bibitem[{Darmanis et~al.(2015)Darmanis, Sloan, Zhang, Enge, Caneda, Shuer,
  Hayden~Gephart, Barres, and Quake}]{darmanis2015survey}
Darmanis, S., Sloan, S.~A., Zhang, Y., Enge, M., Caneda, C., Shuer, L.~M.,
  Hayden~Gephart, M.~G., Barres, B.~A., and Quake, S.~R. (2015), \enquote{A
  survey of human brain transcriptome diversity at the single cell level,}
  \textit{Proceedings of the National Academy of Sciences}, 112, 7285--7290.

\bibitem[{Datta and Zou(2017)}]{datta2017cocolasso}
Datta, A. and Zou, H. (2017), \enquote{Cocolasso for high-dimensional
  error-in-variables regression,} \textit{The Annals of Statistics}, 45,
  2400--2426.

\bibitem[{Erdmann-Pham et~al.(2021)Erdmann-Pham, Fischer, Hong, and
  Song}]{erdmann2021likelihood}
Erdmann-Pham, D.~D., Fischer, J., Hong, J., and Song, Y.~S. (2021),
  \enquote{Likelihood-based deconvolution of bulk gene expression data using
  single-cell references,} \textit{Genome Research}, 31, 1794--1806.

\bibitem[{Fan and Li(2001)}]{fan2001variable}
Fan, J. and Li, R. (2001), \enquote{Variable selection via nonconcave penalized
  likelihood and its oracle properties,} \textit{Journal of the American
  statistical Association}, 96, 1348--1360.

\bibitem[{Goldfarb and Idnani(1982)}]{goldfarb1982dual}
Goldfarb, D. and Idnani, A. (1982), \enquote{Dual and primal-dual methods for
  solving strictly convex quadratic programs,} \textit{Numerical analysis},
  226--239.

\bibitem[{Goldfarb and Idnani(1983)}]{goldfarb1983numerically}
--- (1983), \enquote{A numerically stable dual method for solving strictly
  convex quadratic programs,} \textit{Mathematical programming}, 27, 1--33.

\bibitem[{Greene(2003)}]{greene2003econometric}
Greene, W.~H. (2003), \textit{Econometric analysis}, Pearson Education India.

\bibitem[{Hekselman and Yeger-Lotem(2020)}]{hekselman2020mechanisms}
Hekselman, I. and Yeger-Lotem, E. (2020), \enquote{Mechanisms of tissue and
  cell-type specificity in heritable traits and diseases,} \textit{Nature
  Reviews Genetics}, 21, 137--150.

\bibitem[{Higham(1988)}]{higham1988computing}
Higham, N.~J. (1988), \enquote{Computing a nearest symmetric positive
  semidefinite matrix,} \textit{Linear algebra and its applications}, 103,
  103--118.

\bibitem[{Jaakkola and Elo(2022)}]{jaakkola2022estimating}
Jaakkola, M.~K. and Elo, L.~L. (2022), \enquote{Estimating cell type-specific
  differential expression using deconvolution,} \textit{Briefings in
  bioinformatics}, 23, bbab433.

\bibitem[{Jew et~al.(2020)Jew, Alvarez, Rahmani, Miao, Ko, Garske, Sul,
  Pietil{\"a}inen, Pajukanta, and Halperin}]{jew2020accurate}
Jew, B., Alvarez, M., Rahmani, E., Miao, Z., Ko, A., Garske, K.~M., Sul, J.~H.,
  Pietil{\"a}inen, K.~H., Pajukanta, P., and Halperin, E. (2020),
  \enquote{Accurate estimation of cell composition in bulk expression through
  robust integration of single-cell information,} \textit{Nature
  communications}, 11, 1--11.

\bibitem[{Jin et~al.(2021)Jin, Chen, Lin, and Sun}]{jin2021cell}
Jin, C., Chen, M., Lin, D.-Y., and Sun, W. (2021), \enquote{Cell-type-aware
  analysis of RNA-seq data,} \textit{Nature Computational Science}, 1,
  253--261.

\bibitem[{Krämer et~al.(2013)Krämer, Green, Pollard, and Tugendreich}]{ipa}
Krämer, A., Green, J., Pollard, Jack, J., and Tugendreich, S. (2013),
  \enquote{{Causal analysis approaches in Ingenuity Pathway Analysis},}
  \textit{Bioinformatics}, 30, 523--530.

\bibitem[{Li et~al.(2016)Li, Severson, Pignon, Zhao, Li, Novak, Jiang, Shen,
  Aster, Rodig, et~al.}]{li2016comprehensive}
Li, B., Severson, E., Pignon, J.-C., Zhao, H., Li, T., Novak, J., Jiang, P.,
  Shen, H., Aster, J.~C., Rodig, S., et~al. (2016), \enquote{Comprehensive
  analyses of tumor immunity: implications for cancer immunotherapy,}
  \textit{Genome biology}, 17, 1--16.

\bibitem[{Li et~al.(2003)Li, Wang, Wang, Quon, Liu, and
  Cordell}]{li2003positive}
Li, Y., Wang, H., Wang, S., Quon, D., Liu, Y.-W., and Cordell, B. (2003),
  \enquote{Positive and negative regulation of APP amyloidogenesis by
  sumoylation,} \textit{Proceedings of the National Academy of Sciences}, 100,
  259--264.

\bibitem[{Little et~al.(2022)Little, Zhabotynsky, Li, Lin, and
  Sun}]{little2022cell}
Little, P., Zhabotynsky, V., Li, Y., Lin, D., and Sun, W. (2022), \enquote{Cell
  type-specific Expression Quantitative Trait Loci,} \textit{bioRxiv}.

\bibitem[{Martin et~al.(2007)Martin, Wilkinson, Nishimune, and
  Henley}]{martin2007emerging}
Martin, S., Wilkinson, K.~A., Nishimune, A., and Henley, J.~M. (2007),
  \enquote{Emerging extranuclear roles of protein SUMOylation in neuronal
  function and dysfunction,} \textit{Nature Reviews Neuroscience}, 8, 948--959.

\bibitem[{Mathys et~al.(2019)Mathys, Davila-Velderrain, Peng, Gao, Mohammadi,
  Young, Menon, He, Abdurrob, Jiang, et~al.}]{mathys2019single}
Mathys, H., Davila-Velderrain, J., Peng, Z., Gao, F., Mohammadi, S., Young,
  J.~Z., Menon, M., He, L., Abdurrob, F., Jiang, X., et~al. (2019),
  \enquote{Single-cell transcriptomic analysis of Alzheimer’s disease,}
  \textit{Nature}, 570, 332--337.

\bibitem[{Montembeault et~al.(2016)Montembeault, Rouleau, Provost, and
  Brambati}]{montembeault2016altered}
Montembeault, M., Rouleau, I., Provost, J.-S., and Brambati, S.~M. (2016),
  \enquote{Altered gray matter structural covariance networks in early stages
  of Alzheimer's disease,} \textit{Cerebral cortex}, 26, 2650--2662.

\bibitem[{Mostafavi et~al.(2018)Mostafavi, Gaiteri, Sullivan, White, Tasaki,
  Xu, Taga, Klein, Patrick, Komashko, et~al.}]{mostafavi2018molecular}
Mostafavi, S., Gaiteri, C., Sullivan, S.~E., White, C.~C., Tasaki, S., Xu, J.,
  Taga, M., Klein, H.-U., Patrick, E., Komashko, V., et~al. (2018), \enquote{A
  molecular network of the aging human brain provides insights into the
  pathology and cognitive decline of Alzheimer’s disease,} \textit{Nature
  neuroscience}, 21, 811--819.

\bibitem[{Newman et~al.(2015)Newman, Liu, Green, Gentles, Feng, Xu, Hoang,
  Diehn, and Alizadeh}]{newman2015robust}
Newman, A.~M., Liu, C.~L., Green, M.~R., Gentles, A.~J., Feng, W., Xu, Y.,
  Hoang, C.~D., Diehn, M., and Alizadeh, A.~A. (2015), \enquote{Robust
  enumeration of cell subsets from tissue expression profiles,} \textit{Nature
  methods}, 12, 453--457.

\bibitem[{Newman et~al.(2019)Newman, Steen, Liu, Gentles, Chaudhuri, Scherer,
  Khodadoust, Esfahani, Luca, Steiner, et~al.}]{newman2019determining}
Newman, A.~M., Steen, C.~B., Liu, C.~L., Gentles, A.~J., Chaudhuri, A.~A.,
  Scherer, F., Khodadoust, M.~S., Esfahani, M.~S., Luca, B.~A., Steiner, D.,
  et~al. (2019), \enquote{Determining cell type abundance and expression from
  bulk tissues with digital cytometry,} \textit{Nature biotechnology}, 37,
  773--782.

\bibitem[{Patel et~al.(2021)Patel, Zhang, Farrell, Chung, Stein, Lunetta, and
  Farrer}]{patel2021cell}
Patel, D., Zhang, X., Farrell, J.~J., Chung, J., Stein, T.~D., Lunetta, K.~L.,
  and Farrer, L.~A. (2021), \enquote{Cell-type-specific expression quantitative
  trait loci associated with Alzheimer disease in blood and brain tissue,}
  \textit{Translational Psychiatry}, 11, 1--17.

\bibitem[{Patrick et~al.(2020)Patrick, Taga, Ergun, Ng, Casazza, Cimpean, Yung,
  Schneider, Bennett, Gaiteri, et~al.}]{patrick2020deconvolving}
Patrick, E., Taga, M., Ergun, A., Ng, B., Casazza, W., Cimpean, M., Yung, C.,
  Schneider, J.~A., Bennett, D.~A., Gaiteri, C., et~al. (2020),
  \enquote{Deconvolving the contributions of cell-type heterogeneity on
  cortical gene expression,} \textit{PLoS Computational Biology}, 16, e1008120.

\bibitem[{Reitz et~al.(2011)Reitz, Brayne, and Mayeux}]{reitz2011epidemiology}
Reitz, C., Brayne, C., and Mayeux, R. (2011), \enquote{Epidemiology of
  Alzheimer disease,} \textit{Nature Reviews Neurology}, 7, 137--152.

\bibitem[{Robinson et~al.(2010)Robinson, McCarthy, and
  Smyth}]{robinson2010edger}
Robinson, M.~D., McCarthy, D.~J., and Smyth, G.~K. (2010), \enquote{edgeR: a
  Bioconductor package for differential expression analysis of digital gene
  expression data,} \textit{Bioinformatics}, 26, 139--140.

\bibitem[{Rothman et~al.(2009)Rothman, Levina, and
  Zhu}]{rothman2009generalized}
Rothman, A.~J., Levina, E., and Zhu, J. (2009), \enquote{Generalized
  thresholding of large covariance matrices,} \textit{Journal of the American
  Statistical Association}, 104, 177--186.

\bibitem[{Salat et~al.(2001)Salat, Kaye, and Janowsky}]{salat2001selective}
Salat, D.~H., Kaye, J.~A., and Janowsky, J.~S. (2001), \enquote{Selective
  preservation and degeneration within the prefrontal cortex in aging and
  Alzheimer disease,} \textit{Archives of neurology}, 58, 1403--1408.

\bibitem[{Su et~al.(2021)Su, Zhang, and Zhao}]{su2021csnet}
Su, C., Zhang, J., and Zhao, H. (2021), \enquote{CSNet: Estimating
  cell-type-specific gene co-expression networks from bulk gene expression
  data,} \textit{bioRxiv}.

\bibitem[{Tang et~al.(2020)Tang, Park, and Zhao}]{tang2020nitumid}
Tang, D., Park, S., and Zhao, H. (2020), \enquote{NITUMID: nonnegative matrix
  factorization-based immune-TUmor MIcroenvironment Deconvolution,}
  \textit{Bioinformatics}, 36, 1344--1350.

\bibitem[{Tang et~al.(2022)Tang, Park, and Zhao}]{tang2022scadie}
--- (2022), \enquote{SCADIE: simultaneous estimation of cell type proportions
  and cell type-specific gene expressions using SCAD-based iterative estimating
  procedure,} \textit{Genome biology}, 23, 1--23.

\bibitem[{Tian et~al.(2021)Tian, Wang, and Roeder}]{tian2021esco}
Tian, J., Wang, J., and Roeder, K. (2021), \enquote{ESCO: single cell
  expression simulation incorporating gene co-expression,}
  \textit{Bioinformatics}, 37, 2374--2381.

\bibitem[{Trapnell et~al.(2012)Trapnell, Roberts, Goff, Pertea, Kim, Kelley,
  Pimentel, Salzberg, Rinn, and Pachter}]{trapnell2012differential}
Trapnell, C., Roberts, A., Goff, L., Pertea, G., Kim, D., Kelley, D.~R.,
  Pimentel, H., Salzberg, S.~L., Rinn, J.~L., and Pachter, L. (2012),
  \enquote{Differential gene and transcript expression analysis of RNA-seq
  experiments with TopHat and Cufflinks,} \textit{Nature protocols}, 7,
  562--578.

\bibitem[{Trapnell et~al.(2010)Trapnell, Williams, Pertea, Mortazavi, Kwan,
  Van~Baren, Salzberg, Wold, and Pachter}]{trapnell2010transcript}
Trapnell, C., Williams, B.~A., Pertea, G., Mortazavi, A., Kwan, G., Van~Baren,
  M.~J., Salzberg, S.~L., Wold, B.~J., and Pachter, L. (2010),
  \enquote{Transcript assembly and quantification by RNA-Seq reveals
  unannotated transcripts and isoform switching during cell differentiation,}
  \textit{Nature biotechnology}, 28, 511--515.

\bibitem[{Vershynin(2018)}]{vershynin2018high}
Vershynin, R. (2018), \textit{High-dimensional probability: An introduction
  with applications in data science}, vol.~47, Cambridge university press.

\bibitem[{Wang et~al.(2021)Wang, Roeder, and Devlin}]{wang2021bayesian}
Wang, J., Roeder, K., and Devlin, B. (2021), \enquote{Bayesian estimation of
  cell type-specific gene expression with prior derived from single-cell data,}
  \textit{Genome Research}, gr--268722.

\bibitem[{Wang et~al.(2019)Wang, Park, Susztak, Zhang, and Li}]{wang2019bulk}
Wang, X., Park, J., Susztak, K., Zhang, N.~R., and Li, M. (2019), \enquote{Bulk
  tissue cell type deconvolution with multi-subject single-cell expression
  reference,} \textit{Nature communications}, 10, 1--9.

\bibitem[{Xie and Wang(2022)}]{xie2022robust}
Xie, D. and Wang, J. (2022), \enquote{Robust Statistical Inference for Cell
  Type Deconvolution,} \textit{arXiv preprint arXiv:2202.06420}.

\bibitem[{Yang et~al.(2021)Yang, Alessandri-Haber, Fury, Schaner, Breese,
  LaCroix-Fralish, Kim, Adler, Macdonald, Atwal, et~al.}]{yang2021adroit}
Yang, T., Alessandri-Haber, N., Fury, W., Schaner, M., Breese, R.,
  LaCroix-Fralish, M., Kim, J., Adler, C., Macdonald, L.~E., Atwal, G.~S.,
  et~al. (2021), \enquote{AdRoit is an accurate and robust method to infer
  complex transcriptome composition,} \textit{Communications biology}, 4,
  1--14.

\bibitem[{Yussof et~al.(2020)Yussof, Yoon, Krkljes, Schweinberg, Cottrell, Chu,
  and Chang}]{adpathways}
Yussof, A., Yoon, P., Krkljes, C., Schweinberg, S., Cottrell, J., Chu, T., and
  Chang, S.~L. (2020), \enquote{A meta-analysis of the effect of binge drinking
  on the oral microbiome and its relation to {Alzheimer}’s disease,}
  \textit{Scientific Reports}, 10, 19872.

\bibitem[{Zhang et~al.(2013)Zhang, Gaiteri, Bodea, Wang, McElwee,
  Podtelezhnikov, Zhang, Xie, Tran, Dobrin, et~al.}]{zhang2013integrated}
Zhang, B., Gaiteri, C., Bodea, L.-G., Wang, Z., McElwee, J., Podtelezhnikov,
  A.~A., Zhang, C., Xie, T., Tran, L., Dobrin, R., et~al. (2013),
  \enquote{Integrated systems approach identifies genetic nodes and networks in
  late-onset Alzheimer’s disease,} \textit{Cell}, 153, 707--720.

\bibitem[{Zhang and Horvath(2005)}]{zhang2005general}
Zhang, B. and Horvath, S. (2005), \enquote{A general framework for weighted
  gene co-expression network analysis,} \textit{Statistical applications in
  genetics and molecular biology}, 4.

\bibitem[{Zhang and Li(2022{\natexlab{a}})}]{zhang2022high}
Zhang, J. and Li, Y. (2022{\natexlab{a}}), \enquote{High-Dimensional Gaussian
  Graphical Regression Models with Covariates,} \textit{Journal of the American
  Statistical Association}, 1--13.

\bibitem[{Zhang and Li(2022{\natexlab{b}})}]{zhang2022multi}
--- (2022{\natexlab{b}}), \enquote{Multi-task Learning for Gaussian Graphical
  Regressions with High Dimensional Covariates,} \textit{arXiv preprint
  arXiv:2205.10672}.

\end{thebibliography}
\endgroup

\newpage
\renewcommand{\thesection}{A}
\renewcommand{\thesubsection}{A\arabic{subsection}}
\renewcommand{\theequation}{S\arabic{equation}}
\renewcommand{\thelemma}{S\arabic{lemma}}
\setcounter{equation}{0}
\setcounter{lemma}{0}
\setcounter{table}{0}
\setcounter{figure}{0}
\setcounter{section}{0}
\setcounter{subsection}{0}  
\setcounter{page}{1}
\def\eop
{\hfill $\Box$
}

\baselineskip=21.5pt
\begin{center}
{\large\bf Supplementary Materials for ``Statistical Inference of Cell-type Proportions Estimated from Bulk Expression Data"} \\
\bigskip
\end{center}
\renewcommand{\thetable}{S\arabic{table}}
\renewcommand{\thefigure}{S\arabic{figure}}

\subsection{Additional Computational Details and Results}
\subsubsection{DECALS with sparse \texorpdfstring{$\bSigma^{(k)}$'s}{TEXT}}\label{sec:sparse}
The accumulated errors across $O(p^2)$ entries in the estimated $\bSigma^{(k)}$ can be excessive when $p$, the number of signature genes, much exceeds $n$, the number of bulk samples. In this case, we consider a sparse estimation of ${\bSigma}^{(k)}$.

Let $\D^{(k)}=\text{diag}(\bSigma_{11}^{(k)},\ldots,\bSigma_{pp}^{(k)})$, the correlation matrix $\R^{(k)}$ be $\R_{jj'}^{(k)}=\bSigma_{jj'}^{(k)}/\sqrt{\D_{jj}^{(k)}\D_{j'j'}^{(k)}}$ and denote $\mathcal{T}_{\lambda_k}(\hat{\R}^{(k)})$ as the element-wise SCAD \citep{fan2001variable} thresholded $\R^{(k)}$, calculated as
\begin{equation}\label{scad}
\left[\mathcal{T}_{\lambda_k}({\R}^{(k)})\right]_{jj'}=\text{sign}({\R}^{(k)}_{jj'})\times\left\{
\begin{array}{ll}
\max\{|{\R}^{(k)}_{jj'}|-\lambda_k,0\}, & |{\R}^{(k)}_{jj'}|\leq 2\lambda_k,\\
\lambda_k+\frac{a-1}{a-2}(|{\R}^{(k)}_{jj'}|-2\lambda_k), & |{\R}^{(k)}_{jj'}|\in (2\lambda_k,a\lambda_k),\\
|{\R}^{(k)}_{jj'}|, & \text{otherwise,}
\end{array}
\right.
\end{equation}
where $\lambda_k$ is a tuning parameter that can be selected by cross validation. The parameter $a$ is set as 3.7 as recommended by \cite{fan2001variable}. 

\begin{algorithm}
\begin{algorithmic}
\caption{\texttt{DECALS} with sparse $\bSigma^{(k)}$'s}
\STATE \textbf{Input:} Bulk expressions $\{\bm{y}_i\}_{1\le i\le n}$ and the signature gene matrix $\bm{W}$.\\
\STATE \hspace{0.25in} \textbf{Step 1:} Calculate the constrained least squares estimator $\hat{\bpi}_i$ from \eqref{opt} for $1\le i\le n$.
\STATE \hspace{0.25in} \textbf{Step 2:} Initialize $\V_i^{[0]}$ for $1\le i\le n$ and tuning $\lambda_k$'s via cross validation.
\STATE \hspace{0.25in} \textbf{Repeat} the following steps for $t=0,1,\ldots$ until convergence.
\STATE \hspace{0.5in} \textbf{Step 3.1:} Calculate $\bm{B}^{[t]}_1$ with \eqref{firstest}, $\hat\bpi_i$ and $\V_i^{[t]}$.
\STATE \hspace{0.5in} \textbf{Step 3.2:} Calculate $(\bSigma^{(k)})^{[t]}$ with \eqref{sigmakest1}, $\hat\H$, $\hat\z_j$ and $\bm{B}^{[t]}_1$.
\STATE \hspace{0.5in} \textbf{Step 3.3:} Find $(\R^{(k)})^{[t]}$ and set $(\check\bSigma^{(k)})^{[t]}=(\D^{(k)})^{[t]}\mathcal{T}_{\lambda_k}((\R^{(k)})^{[t]})(\D^{(k)})^{[t]}$.  \STATE \hspace{0.5in} \textbf{Step 3.4:} Calculate $(\bSigma^{(k)})^{[t]}=\arg\min_{\bSigma\succ 0}\|\bSigma-(\check\bSigma^{(k)})^{[t]}\|_F$.
\STATE \hspace{0.5in} \textbf{Step 3.5:} Calculate $\bSigma_i^{[t]}$ with \eqref{ass1eq}, $\hat\bpi_i$ and $(\bSigma^{(k)})^{[t]}$.
\STATE \hspace{0.5in} \textbf{Step 3.6:} Calculate $\V_i^{[t+1]}$ with \eqref{asydist}, $\W$ and $\bSigma_i^{[t]}$.
\STATE \textbf{Output:} The estimated proportions $\{\hat\bpi_i\}_{1\le i\le n}$ and covariances $\{\hat\V_i\}_{1\le i\le n}$.
\end{algorithmic}
\end{algorithm}
The parameters $\lambda_k$'s in Step 2 are tuned using cross validation with $\hat{\bpi}_i$ and $\{\bm{y}_i\}_{1\le i\le n}$ as in \citet{su2021csnet}.
In step 3.4, $(\bSigma^{(k)})^{[t]}$ is calculated as the nearest symmetric and positive semi-definite matrix in the Frobenius norm to $(\check\bSigma^{(k)})^{[t]}$. This is a commonly adopted procedure \citep{chen2009similarity,bradley2015multivariate,datta2017cocolasso} and can be efficiently calculated using \citet{higham1988computing}.

\subsubsection{Estimating \texorpdfstring{$\hat{\bpi}_i^{\text{GLS}}$}{TEXT} and \texorpdfstring{$\hat{\V}_i^{\text{GLS}}$}{TEXT}}\label{gls}
Given $\bSigma_i$, we estimate $\hat{\bpi}_i^{\text{GLS}}$ via the following constrained generalized least squares (GLS):
$$
\begin{aligned}
&\min_{\bpi_i\in\mathbb{R}^{K}}\left\Vert\bSigma_i^{-1/2}\y_i-\bSigma_i^{-1/2}\W\bpi_i\right\Vert_2^2,\\
&\text{s.t.}\,\,\pi_{ik}\geq 0 \text{  and  } \sum_{k=1}^K\pi_{ik}=1,
\end{aligned}
$$
and $\V^{\text{GLS}}_i$ via \eqref{glsvar}, which is
$$
\V^{\text{GLS}}_i=
(\W^\top\bSigma^{-1}_i\W)^{-1}\left\{\I-\1\{\1^\top(\W^\top\bSigma^{-1}_i\W)^{-1}\1^\top\}^{-1}\1^\top(\W^\top\bSigma^{-1}_i\W)^{-1}\right\}.
$$
We iteratively update $\bpi_i^{\text{GLS}}$, $\bSigma_i$ and $\V^{\text{GLS}}_i$ in the estimation procedure, as detailed in the following algorithm.

\begin{algorithm}
\caption{\bf The constrained generalized least squares algorithm.}
\begin{algorithmic}
\STATE \textbf{Input:} Bulk expression $\{\bm{y}_i\}_{1\leq i\leq n}$ and the signature gene matrix $\bW$.
\STATE \textbf{Step 1:} Initialize $\bSigma_i^{[0]}=\I$.
\STATE \textbf{Repeat} the following steps for $t=0,1,...$ until convergence.
\STATE \hspace{0.3in} \textbf{Step 2.1:} Calculate $(\bpi_i^{\text{GLS}})^{[t]}$ with \eqref{opt2} and $\bSigma_i^{(t)}$.
\STATE \hspace{0.3in} \textbf{Step 2.2:} Calculate $(\bSigma^{(k)})^{[t]}$ with \eqref{sigmakest0} and $(\bpi_i^{\text{GLS}})^{[t]}$.
\STATE \hspace{0.3in} \textbf{Step 2.3} Calculate $\bSigma_i^{[t]}$ with \eqref{ass1eq}, $(\bpi_i^{\text{GLS}})^{[t]}$ and $(\bSigma^{(k)})^{[t]}$.
\STATE \hspace{0.3in} \textbf{Step 2.4} Calculate $(\V_i^{\text{GLS}})^{[t]}$ with \eqref{glsvar}, $\bW$ and $\bSigma_i^{[t]}$.
\STATE \textbf{Output:} The estimated proportion $\{\hat{\bpi}_i^{\text{GLS}}\}_{1\leq i\leq n}$ and covariance $\{\hat{\V}_i^{\text{GLS}}\}_{1\leq i\leq n}$.
\end{algorithmic}
\label{alg:gls}
\end{algorithm}

\subsubsection{Estimation accuracy of \texorpdfstring{$\V_i$}{TEXT}} 
We investigate the estimation accuracy with different numbers of genes $p=150, 300$ and different signal strengths in the signature matrix. 
When $p=150$, we consider similar covariances as in Figure \ref{fig5}, with block sizes replaced by 50. We generate $w_{kj}\overset{i.i.d.}{\sim} \mathcal{N}(0,a^2)$ for all $k,j$. 
When $a$ is large, the signature matrix is more ``informative" and the errors in estimating $\V_i$ are expected to decrease.
The estimation errors of $\V_i$, calculated as $\sqrt{\frac{1}{n}\sum_{i=1}^n\|\hat{\V}_{i,ll'}-\V_{i,ll'}\|_2^2}$, are summarized in Table~\ref{tab1}. It is seen that the estimation accuracy of \texttt{DECALS} improves with $p$, the number of signature genes, and $a$, the variance of signature gene expressions. 
\begin{table}[!t]
	\centering
	{\renewcommand{\arraystretch}{0.85}
{\begin{tabular}{c|c|c|cccccc} \hline
	       \multicolumn{3}{c}{$(l,l')$}\vline  & (1,1) & (2,2) & (3,3) & (1,2) & (1,3) & (2,3)\\ \hline
		\multirow{8}{*}{$p=150$} & \multirow{4}{*}{$a=1$} & \texttt{DECALS} &4.741 &5.019 &6.256 &2.663 &2.832 &3.970\\ 
		 & & {\small [$\times 10^{-3}$]} & (0.045) & (0.033) & (0.075) & (0.019) & (0.046) & (0.029) \\ \cline{3-9}
		&& OLS &19.875 &14.996 &11.233  &4.802 & 7.168 &11.013\\ 
		 & &  {\small [$\times 10^{-3}$]}& (0.054) & (0.053) & (0.050) & (0.001) & (0.001) & (0.001) \\ \cline{2-9}
		 & \multirow{4}{*}{$a=2$} & \texttt{DECALS} &0.811 &1.501 &1.756 &0.356 &0.534 &1.251\\ 
		 & &  {\small [$\times 10^{-3}$]}& (0.018) & (0.024) & (0.024) & (0.011) & (0.012) & (0.018) \\ \cline{3-9}
		&& OLS &3.890 &4.686 &3.854 &1.878 &2.629 &3.557\\ 
		 & &  {\small [$\times 10^{-3}$]}& (0.012) & (0.016) & (0.015) & (0.001) & (0.001) & (0.000) \\ \hline
		 \multirow{8}{*}{$p=300$} & \multirow{4}{*}{$a=1$} & \texttt{DECALS} &1.475 &1.692 &1.866 &0.749 &0.840 &1.175\\ 
		 & &  {\small [$\times 10^{-3}$]}& (0.015) & (0.016) & (0.018) & (0.013) & (0.011) & (0.013) \\ \cline{3-9}
		&& OLS &10.104  &7.534 & 9.205  &3.884 & 5.733 & 7.697\\ 
		 & &  {\small [$\times 10^{-3}$]}& (0.022) & (0.021) & (0.023) & (0.000) & (0.002) & (0.002) \\ \cline{2-9}
		 & \multirow{4}{*}{$a=2$} & \texttt{DECALS} &0.324 &0.496 &0.735 &0.117 &0.282 &0.458\\ 
		 & &  {\small [$\times 10^{-3}$]}& (0.006) & (0.008) & (0.009) & (0.003) & (0.006) & (0.006) \\ \cline{3-9}
		&& OLS &1.882 &1.859 &1.456 &1.089 &1.147 &1.586\\ 
		 & &  {\small [$\times 10^{-3}$]}& (0.004) & (0.005) & (0.005) & (0.000) & (0.000) & (0.000) \\ \hline
	\end{tabular}}}
	\caption{The estimation errors of $\V_{i,ll'}$ when $p=150,300$ and $a=1,2$. }
	\label{tab1}
\end{table}

\subsubsection{Simulation in Section \ref{set2} with Gaussian distributions}\label{normal}
\begin{figure}[!t]
\centering
\includegraphics[trim=0 10mm 0 0, scale=0.75]{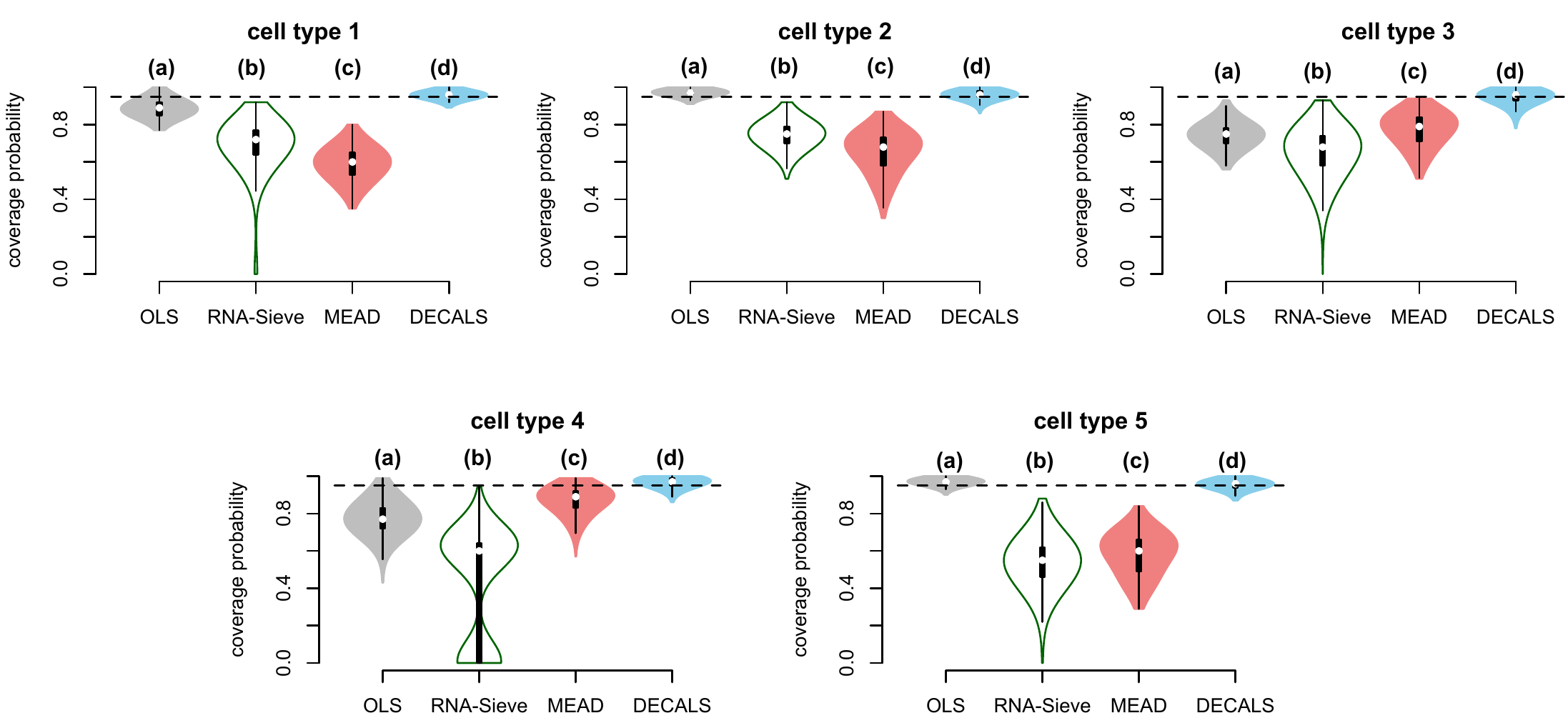}
\caption{\label{result2} The coverage probabilities of 95\% confidence intervals in Gaussian data with (a) \texttt{OLS}, (b) \texttt{RNA-Sieve}, (c) \texttt{MEAD} and (d) \texttt{DECALS}.}
\end{figure}

We consider the case where the CTS expression profiles $x_{ij}^{(k)}$'s are Gaussian.
Similar as Section \ref{set1}, the bulk gene expression for sample $i$ is calculated as $\y_i=\sum_{k=1}^K\pi_{ik}\x_i^{(k)}$, where the expression profile $\x_i^{(k)}$ is simulated from $\x_i^{(k)}\sim\mathcal{N}(\w_k,\bSigma^{(k)})$. 
We apply \texttt{OLS}, \texttt{MEAD}, \texttt{RNA-Sieve} and \texttt{DECALS} to infer cell-type proportions for each subject. Specifically, we construct 95\% confidence intervals for $\pi_{ik}$'s using each method and estimate the coverage probabilities using 100 data replicates. 
The results are summarized in Figure \ref{result2}.
It is seen that DECALS has the best performance in all five cell types, with coverage probabilities close to the nominal level of 95\%. 


\subsubsection{Simulating from a joint distribution with Gamma marginals}\label{sec::gamma}


We consider a copula method that is similar to \citet{tian2021esco}. 
Specifically, we generate the CTS expression of gene $j$ in cell type $k$ by 
$x^{(k)}_{j}=F_{kj}^{-1}(\Phi(z^{(k)}_{j}))$ where $(z^{(k)}_{1},..., z^{(k)}_{p}) \sim \mathcal{N}(0,\R^{(k)})$, $\R^{(k)}$ is the target correlation matrix, $\Phi$ is the cumulative distribution function (CDF) of the standard normal, and $F_{kj}^{-1}$ is the inverse CDF of a Gamma distribution with shape parameter $\alpha_{kj}$ and scale parameter $\theta_{kj}$. 
We set $\alpha_{kj}=0.01$ and $\theta_{kj}=w_{kj}/\alpha_{kj}$ to ensure $\alpha\theta_{kj}=w_{kj}$ for all $k$ and $j$. 
Correspondingly, the generated CTS expression $x^{(k)}_{ij}$ follows a Gamma distribution with mean $w_{jk}$.

\subsection{Technical Lemmas}
We state the technical lemmas that will be used in our proofs. 
\begin{lemma}\label{thm1}
Assume that $\bW^\top\bW\in\mathbb{R}^{K\times K}$ is positive definite, all $\pi_{ik}>0$, $\epsilon_{ij}$'s are sub-exponential random variables and $\|\bSigma_i\|_{1,0}<c$ for some positive constant $c$. There exists a sufficiently large $p\geq p_0$ such that the constrained least-squares estimator vector $\hat{\bpi}_i$ from \eqref{opt} reduces to the equality constrained least squares estimate written as
\begin{equation}\label{eclse}
    \hat{\bpi}_{i,eq}=\tilde{\bpi}_i-(\bW^\top\bW)^{-1}\1_k\{\1_k^\top(\bW^\top\bW)^{-1}\1_k\}^{-1}(\1_k^\top\tilde{\bpi}_i-1),
\end{equation}
where $\tilde{\bpi}_i=(\bW^\top\bW)^{-1}\bW^\top\y_i$ is the ordinary least squares estimator.
\end{lemma}

\begin{lemma}[Theorem 1 in \cite{zhang2022multi}]
\label{netinequ} 
Consider $N$ correlated mean zero sub-exponential random variables $Y_j$, $j\in[N]$ and an induced network $G(V, E)$ with a node set $V = {1,\ldots, N}$ and an edge set $E =
\{(j, k) : \Cov(Y_j
, Y_k) \neq 0\}$. Denote the maximum node degree of $G(V, E)$ by $d_{\max}$ and let $c_G=\min\left(d_{\max}+1,\frac{1+\sqrt{8|E|+1}}{2}\right)$. For any $t\geq 0$ and a constant $c > 0$, it holds that
\begin{equation*}
    \mathbb{P}\left(\sum_{j=1}^N Y_j\geq t\right)\leq c_G\exp\left[-c\min\left\{\frac{t^2}{c_G^2\sum_{j=1}^N\|Y_j\|_{\psi_1}^2},\frac{t}{c_G\max_j\|Y_j\|_{\psi_1}}\right\}\right],
\end{equation*}
where $\|\cdot\|_{\psi_1}$ is the sub-exponential norm.
\end{lemma}

\begin{lemma} [Proposition 2.7.1 in \cite{vershynin2018high}]
\label{subexpmom}
For a mean zero sub-exponential random variable $X$, its moments satisfy 
$(\mathbb{E}|X|^p)^{1/p}\leq c_1 p$, for all $p\geq 1$ and some positive constant $c_1$.
\end{lemma}

\begin{lemma}[Theorem 2.7 in \cite{chen2004normal}] 
\label{dependantclt}
Let $\{X_i,i\in\mathcal{V}\}$ be random variables indexed by the vertices of a dependency graph, $g(\mathcal{V},\mathcal{E})$ and let $D$ be the maximum degree. Put $W=\sum_{i\in\mathcal{V}}X_i$. Assume that $\mathbb{E}W^2=1$, $\mathbb{E}X_i = 0$ and $\mathbb{E}|X_i|^{p}\leq \theta^p$ for $i\in \mathcal{V}$ and some $\theta>0$. Then
\begin{equation*}
    \sup_{z}|\mathbb{P}(W\leq z)-\phi(z)|\leq 75 D^{5(p-1)}|\mathcal{V}|\theta^p.
\end{equation*}
\end{lemma}

\subsection{Proof of Lemma~\ref{thm1}}
The constraints in \eqref{opt} can be written as $\1_K^\top\bpi_i=1$ and $\I_K\bpi_i\succeq \0$, where $\a\succeq\b$ means $a_{j}>b_j$ for all $j$. These constraints can be rewritten as 
\begin{equation}
    \label{constraints}
    \begin{aligned}
    &\I_K\bpi_i-\v_1=\0,\\
    &\1_K^\top\bpi_i=1,
    \end{aligned}
\end{equation}
where $\v_1\in\mathbb{R}^{K}$ is a surplus vector.
By \eqref{dual}, the dual function is given by 
\begin{equation*}
    \begin{aligned}
    &\max_{\blambda} \c^\top\blambda+\frac{1}{2}(\y_i^\top\y_i-\bpi_i^\top\bW^\top\bW\bpi_i),\\
    &\text{s.t.}\,\,\A^\top\blambda+\bW^\top\y_i=(\bW^\top\bW)\bpi_i.
    \end{aligned}
\end{equation*}
Let $\blambda=(\blambda_1^\top,\lambda_2)^\top$, where $\blambda_1\in\mathbb{R}^K$ and $\lambda_2\in\mathbb{R}$ are the dual vector and scalar for the inequality and equality constraints, respectively. The constraint of the dual function can be reorganized as
\begin{equation}\label{dualcons}
(\bW^\top\bW)^{-1}\blambda_1+(\bW^\top\bW)^{-1}\1_K\lambda_2+(\bW^\top\bW)^{-1}\bW^\top\y_i=\bpi_i.
\end{equation}
By the Dantzig-Cottle conditions \citep{dantzig1968complementary}, we have
\begin{equation}\label{dualpara}
    \v_1^\top\blambda_1=0,\,\,\v_1\succeq 0,\,\,\blambda_1\succeq 0\,\,\text{and}\,\, \lambda_2\geq 0.
\end{equation}
Let $\M_{11}=(\bW^\top\bW)^{-1}$, $\M_{12}=\M_{12}^\top=(\bW^\top\bW)^{-1}\1_K$ and $\M_{22}=\1_K^\top(\bW^\top\bW)^{-1}\1_K$. Multiplying $\1_K^\top$ to the left and right sides of \eqref{dualcons}, we can get that
\begin{equation}\label{lambda2}
    \lambda_2=-\M_{22}^{-1}\M_{21}\blambda_1-\M_{22}^{-1}(\1_K^\top\tilde\bpi_i-1),
\end{equation}
where $\tilde\bpi_i=(\bW^\top\bW)^{-1}\bW^\top\y_i$.

By \eqref{constraints} and \eqref{dualcons}, we can get that
\begin{equation}\label{v1part}
    \v_1=\I_K\bpi_i=\M_{11}\blambda_1+\M_{12}\lambda_2+\I_K\tilde\bpi_i.
\end{equation}
Plugging \eqref{lambda2} into \eqref{v1part}, it arrives at
\begin{equation}\label{v1}
\begin{aligned}
    \v_1&=\M_{11}\blambda_1-\M_{12}\left\{\M_{22}^{-1}\M_{21}\blambda_1+\M_{22}^{-1}(\1_K^\top\hat\bpi_i-1)\right\}+\I_K\tilde\bpi_i\\
    &=\underbrace{(\M_{11}-\M_{12}\M_{22}^{-1}\M_{21})}_{\T_1}\blambda_1+\underbrace{\tilde\bpi_i-\M_{12}\M_{22}^{-1}(\1_K^\top\tilde\bpi_i-1)}_{\T_2}.
\end{aligned}    
\end{equation}
Since $\tilde\bpi_i=(\bW^\top\bW)^{-1}\bW^\top\y_i=\bpi_i+(\bW^\top\bW)^{-1}\bW^\top\bm{\epsilon}_i$, the term $\T_2$ can be rewritten as
\begin{equation*}
    \T_2=\bpi_i-\underbrace{\M_{12}\M_{22}^{-1}(\1_K^\top\bpi_i-1)}_{\T_{21}}+\underbrace{(\I_K-\M_{12}\M_{22}^{-1}\1_K^\top)\left(\frac{1}{p}\bW^\top\bW\right)^{-1}\frac{1}{p}\bW^\top\bm{\epsilon}_i}_{\T_{22}}.
\end{equation*}
By \eqref{constraints}, it is easily seen that $\T_{21}=\0$. 
Noting $c_G=\|\bSigma_i\|_{1,0}=O(1)$ and applying Lemma~\ref{netinequ}, we have, for any $e>0$,
\begin{equation}\label{plimit}
    \lim_{p\rightarrow\infty}\mathbb{P}\left(\left|\frac{1}{p}\sum_{j=1}^pw_{kj}\epsilon_{ij}\right|>e\right)=0.
\end{equation}
Since all $\pi_{ik}>0$, it is evident that there exists a sufficiently large $p \geq p_0$ such that $\T_2\succ 0$.

Next, we show that $\v_1\succ 0$ and $\blambda_1=\0$. Let
\begin{equation*}
    \M=\begin{pmatrix}
    \M_{11} & \M_{12}\\
    \M_{21} & \M_{22}
    \end{pmatrix}=\begin{pmatrix}
    \I_K\\
    \1_K^\top
    \end{pmatrix}(\bW^\top\bW)^{-1}(\I_K\,\,\1_K).
\end{equation*}
Since $(\bW^\top\bW)^{-1}$ is positive definite, it follows that $\M$ is non-negative definite. We can rewrite $\M$ as
\begin{equation*}
    \M=\begin{pmatrix}
    \M_{11} & \M_{12}\\
    \M_{21} & \M_{22}
    \end{pmatrix}=\begin{pmatrix}
    \I& \M_{12}\M_{22}^{-1}\\
    \0 & \I
    \end{pmatrix}\begin{pmatrix}
    \M_{11}-\M_{12}\M_{22}^{-1}\M_{21} & \0\\
    \0 & \M_{22}
    \end{pmatrix}\begin{pmatrix}
    \I& \0\\
    \M_{22}^{-1}\M_{21} & \I
    \end{pmatrix}.
\end{equation*}
From this, we can conclude that $\B_1=\M_{11}-\M_{12}\M_{22}^{-1}\M_{21}$ is non-negative definite, since $\M$ is non-negative definite and $\M_{22}$ is invertible. By the Dantzig-Cottle conditions in \eqref{dualpara}, we have $\v_1^\top\blambda_1=0$ and then $\blambda_1^\top\B_1\blambda_1+\B_2^\top\blambda_1=0$. Since  $\blambda_1\succeq 0$, we can get that $\blambda_1=\0$. When $\B_2\succ 0$, we can use \eqref{v1} to get that $\v_1=\B_2\succ 0$.
Combined with \eqref{dual} and \eqref{dualcons}, we have, when $p\geq p_0$, 
\begin{equation*}
\begin{aligned}
&\lambda_2=-\M_{22}^{-1}(\1_K^\top\tilde\bpi-1),\\
&\bpi_i=\tilde\bpi_i-(\bW^\top\bW)^{-1}\1_K(\1_K^\top(\bW^\top\bW)^{-1}\1_K)^{-1}(\1_K^\top\tilde\bpi-1).
\end{aligned}   
\end{equation*}
This concludes the proof.

\subsection{Proof of Theorem~\ref{thm2}}\label{sec:prooft1}
We first show the consistency of $\hat\bpi_i$ and then establish its asymptotic normality $\hat\bpi_i$. 
By Lemma~\ref{thm1}, we have that $\hat\bpi_i$ reduces to the equality constrained least-squares estimator $\hat\bpi_{i,eq}$ when $p\geq p_0$. Therefore, $\hat\bpi_i\overset{P}{\rightarrow}\bpi_i$ holds if we can show $\hat\bpi_{i,eq}\overset{P}{\rightarrow}\bpi_i$. Since $\tilde\bpi_i=(\bW^\top\bW)^{-1}\bW^\top\y=\bpi_i+(\bW^\top\bW)^{-1}\bW^\top\bm{\epsilon}_i$, the estimator $\hat\bpi_{i,eq}$ can be rewritten as
\begin{equation*}
    \begin{aligned}
    \hat\bpi_{i,eq}&=\tilde\bpi_i-(\bW^\top\bW)^{-1}\1_K(\1_K^\top(\bW^\top\bW)^{-1}\1_K)^{-1}(\1_K^\top\tilde\bpi_i-1)\\
    &=\bpi_i+\left\{\I_k-(\bW^\top\bW)^{-1}\1_K(\1_K^\top(\bW^\top\bW)^{-1}\1_K)^{-1}\1_K^\top\right\}\left(\frac{1}{p}\bW^\top\bW\right)^{-1}\frac{1}{p}\bW^\top\bm{\epsilon_i}.
    \end{aligned}
\end{equation*}
By \eqref{plimit}, we can get that $$
\left\{\I_k-(\bW^\top\bW)^{-1}\1_K(\1_K^\top(\bW^\top\bW)^{-1}\1_K)^{-1}\1_K^\top\right\}(\bW^\top\bW)^{-1}\bW^\top\bm{\epsilon}_i\overset{P}{\rightarrow} 0.
$$
It then follows that $\hat\bpi_{i,eq}\overset{P}{\rightarrow}\bpi_i$.

\medskip
To derive the asymptotic distribution of $\hat\bpi_i$, we first show the asymptotic distribution of the least squares estimator $\tilde\bpi_i$. After that, we can obtain the asymptotic distribution of the equality constrained least squares estimator $\hat\bpi_{i,eq}$. Finally, the desired result can be obtained by Lemma~\ref{thm1}.

Let $Z_{kj}=\frac{1}{\sqrt{p}}w_{kj}\epsilon_{ij}$ where $w_{kj}$ and $\epsilon_{ij}$ are as defined in \eqref{eqn:de}. As $\epsilon_{ij}$'s are assumed to be mean zero sub-exponential random variables, applying Lemma~\ref{subexpmom} gives
\begin{equation}\label{xkjbound}
    \mathbb{E}(Z_{kj})=0 \text{ and } \mathbb{E}|Z_{kj}|^{2+\sigma}\leq \left(\frac{c_1w_{kj}(2+\sigma)}{\sqrt{p}}\right)^{2+\sigma},
\end{equation}  
for any $\sigma\geq 0$. 
Let $\bm{t}=(t_1,\ldots,t_K)\in\mathbb{R}^K$ be a deterministic vector with $\|\bm{t}\|_2=1$ and define $X_j=\sum_{k=1}^K t_kZ_{kj}$. By \eqref{xkjbound}, it is straightforward to get that $\mathbb{E}(X_j)=0$ and $\mathbb{E}|X_j|^{2+\sigma}=o(p^{-\frac{2+\sigma}{2}})$. 
Moreover, it holds that
\begin{equation*}
\Var\left(\sum_{j=1}^p X_j\right)\rightarrow\bm{t}^\top\bOmega^{-1}\G_i\bOmega^{-1}\bm{t}.
\end{equation*}  
Noting $\|\bSigma_i\|_{1,0}=O(1)$ and by Lemma~\ref{dependantclt}, we have
\begin{equation}
    \sum_{j=1}^p X_j\overset{d}{\rightarrow}N\left(0,\bm{t}^\top\bOmega^{-1}\G_i\bOmega^{-1}\bm{t}\right).
\end{equation}
Next, by Cramer-Wold theorem, it arrives at
\begin{equation}
\frac{1}{\sqrt{p}}\bW^\top\bm{\epsilon}_i=\sum_{j=1}^p\bm{Z}_j\overset{d}{\rightarrow}N\left(0,\bOmega^{-1}\G_{i}\bOmega^{-1}\right).
\end{equation}

Next, letting $\U=\I-\left(\frac{1}{p}\bW^\top\bW\right)^{-1}\1_k\left\{\1_k^\top\left(\frac{1}{p}\bW^\top\bW\right)^{-1}\1_k\right\}^{-1}\1_k^\top$, 
the equation constrained estimator $\hat\bpi_{i,eq}$ can be written as
\begin{equation*}
    \hat{\bpi}_{i,eq}=\U\tilde\bpi_i.
\end{equation*}
Then it is straightforward to get that
\begin{equation}\label{asydist1}
    \sqrt{p}(\hat\bpi_{i,eq}-\bpi_i)\rightarrow \mathcal{N}(\0,\V_i),
\end{equation}
where $\V_i=\U\D\U^\top$, $\D=\left(\frac{1}{p}\bW^\top\bW\right)^{-1}\left(\frac{1}{p}\bW^\top\bSigma_i\bW\right)\left(\frac{1}{p}\bW^\top\bW\right)^{-1}$. 
This combined with Lemma~\ref{thm1} gives the desired conclusion that
\begin{equation}\label{asydist2}
\V_i^{-1/2}\sqrt{p}(\hat\bpi_{i}-\bpi_i)\rightarrow \mathcal{N}(\0,\I).
\end{equation}


\subsection{Proof of Proposition \ref{lem1}}\label{sec:proof2}
For $\B_1$, it is defined as $\mathbb{E}(\hat{\H}^\top\hat{\H})-\H^\top\H$. Let $\hat\bpi_i=\bpi_i+\e_i$ with $\e_i\sim \mathcal{N}(\0,\V_i/p)$, $\mathbb{E}(\hat{\H}^\top\hat{\H})$ can be expanded as
\begin{equation}\label{B1expand}
    \begin{split}
        \mathbb{E}(\hat{\H}^\top\hat{\H})&=\sum_{i=1}^n\mathbb{E}\left[(\bpi_i^{\circ 2}+\e_i^{\circ 2}+2\bpi_i\circ \e_i)(\bpi_i^{\circ 2}+\e_i^{\circ 2}+2\bpi_i\circ \e_i)^\top\right]\\
        &=\sum_{i=1}^n\mathbb{E}\left[(\bpi_i^{\circ 2})(\bpi_i^{\circ 2})^\top\right]+\sum_{i=1}^n\mathbb{E}\left[(\e_i^{\circ 2})(\e_i^{\circ 2})^\top\right]+\sum_{i=1}^n\mathbb{E}\left[4(\bpi_i\circ \e_i)(\bpi_i\circ \e_i)^\top\right]\\
        &\quad+\sum_{i=1}^n\mathbb{E}\left[\bpi_i^{\circ 2}(\e_i^{\circ 2})^\top+(\e_i^{\circ 2})(\bpi_i^{\circ 2})^\top\right]\\
        &\quad+\sum_{i=1}^n\mathbb{E}\left[4(\bpi_i^{\circ 2})(\bpi_i\circ \e_i)^\top\right]+\sum_{i=1}^n\mathbb{E}\left[4(\e_i^{\circ 2})(\bpi_i\circ \e_i)^\top\right].
    \end{split}
\end{equation}
By $\mathbb{E}(\e_i)=\0$, $\mathbb{E}(\e_i^{\circ 2})=\u_i/p$ and $\mathbb{E}(e_{ij}e_{ij'})=V_{i,jj'}/p$, we can get that
\begin{equation}\label{b1term1}
\begin{split}
    &\mathbb{E}\left[(\bpi_i\circ \e_i)(\bpi_i\circ \e_i)^\top\right]=\left\{(\bpi_i)(\bpi_i)^\top\right\}\circ \V_i/p,\\
    &\mathbb{E}\left[\bpi_i^{\circ 2}(\e_i^{\circ 2})^\top+(\e_i^{\circ 2})(\bpi_i^{\circ 2})^\top\right]=(\bpi_i^{\circ 2})\u_i/p+\u_i(\bpi_i^{\circ 2})/p,\\
    &\mathbb{E}\left[(\bpi_i^{\circ 2})(\bpi_i\circ \e_i)^\top\right]=\mathbb{E}\left[(\e_i^{\circ 2})(\bpi_i\circ \e_i)^\top\right]=\0,
\end{split}    
\end{equation}
where $\u_i=(V_{i,11},\ldots,V_{i,KK})^\top$ contain all diagonal terms of $\V_i$. 

The remaining term is $\mathbb{E}\left[(\e_i^{\circ 2})(\e_i^{\circ 2})^\top\right]$. For each element $(j,j')$,  given $e_{ij}\sim \mathcal{N}(0,V_{i,jj}/p)$, $e_{ij'}\sim \mathcal{N}(0,V_{i,j'j'}/p)$ and $\Cov(e_{ij},e_{ij'})=V_{i,jj'}/p$, we have
\begin{equation*}
    \begin{aligned}
    \mathbb{E}(e_{ij}^2e_{ij'}^2)=\left\{\mathbb{E}(e_{ij}e_{ij'})\right\}^2+\Var(e_{ij}e_{ij'}),
    \end{aligned}
\end{equation*}
where for $j\neq j'$
\begin{equation*}
    \mathbb{E}(e_{ij}e_{ij'})=\Cov(e_{ij},e_{ij'})+\mathbb{E}(e_{ij})\mathbb{E}(e_{ij'})=V_{i,jj'}/p,
\end{equation*}
and
\begin{equation*}
    \begin{aligned}
    \Var(e_{ij}e_{ij'})&=\mathbb{E}\left\{\Var(e_{ij}e_{ij'}|e_{ij'})\right\}+\Var\left\{\mathbb{E}(e_{ij}e_{ij'}|e_{ij'})\right\}\\
    &=\mathbb{E}\left\{e_{ij'}^2\Var(e_{ij}|e_{ij'})\right\}+\Var\left\{e_{ij'}\mathbb{E}(e_{ij}|e_{ij'})\right\}\\
    &=\mathbb{E}\left\{e_{ij'}^2(V_{i,jj}-V_{i,jj'}V_{i,j'j'}^{-1}V_{i,j'j})/p\right\}+\Var\left\{e_{ij'}V_{i,jj'}V_{i,j'j'}^{-1}
    e_{ij'}\right\}\\
    &=\frac{1}{p^2}V_{i,j'j'}(V_{i,jj}-V_{i,jj'}V_{i,j'j'}^{-1}V_{i,j'j})+\frac{1}{p^2}V_{i,jj'}^2V_{i,j'j'}^{-2}(2V_{i,j'j'}^{2})\\
    &=\frac{1}{p^2}(V_{i,jj}V_{i,j'j'}+V_{i,jj'}^2).
    \end{aligned}
\end{equation*}
Then it is straightforward to get that
\begin{equation}\label{b1term2}
\begin{aligned}
    \mathbb{E}\left[(e_{ij}^2)(e_{ij'}^2)^\top\right]=\frac{1}{p^2}\left(V_{i,jj}V_{i,j'j'}+2V_{i,jj'}^2\right).
\end{aligned}    
\end{equation}
Additionally, we have $\mathbb{E}\left[e_{ij}^4\right]=\frac{3V_{i,jj}^2}{p^2}$. Thus, we have
\begin{equation*}
    \mathbb{E}\left[(\e_{i}^{\circ 2})(\e_{i}^{\circ 2})^\top\right]=\frac{1}{p^2}\T_{i},
\end{equation*}
where $T_{i,jj'}=V_{i,jj}V_{i,j'j'}+2V_{i,jj'}^2$.
Plugging \eqref{b1term1} and \eqref{b1term2} into \eqref{B1expand}, we get that
\begin{equation*}
\begin{aligned}
    \B_1&=\mathbb{E}(\hat{\H}^\top\hat{\H})-\H^\top\H\\
    &=\frac{1}{p}\sum_i\bpi^{\circ 2}_i \u_i^\top+\frac{1}{p}\sum_i\u_i\bpi^{\circ 2}_i{}^\top+\frac{4}{p}\sum_i(\bpi^{\circ 2}_i{}^\top\bpi^{\circ 2}_i)\circ \V_i+\frac{1}{p^2}\sum_i\T_{i}.
\end{aligned}    
\end{equation*}

For $\B_2$, it is defined as $\mathbb{E}(\hat{\H})-\H$. Similar as $\B_1$, it can be expanded as 
\begin{equation}
\begin{aligned}
    \mathbb{E}(\hat{\H})&=\begin{pmatrix}
    \mathbb{E}(\hat{\bpi}_1^2)^\top\\
    \vdots\\
    \mathbb{E}(\hat{\bpi}_n^2)^\top
    \end{pmatrix}=\begin{pmatrix}
    \mathbb{E}(\bpi_1^{\circ 2}+\e_1^{\circ 2}+2\bpi_1\circ \e_1)^\top\\
    \vdots\\
    \mathbb{E}(\bpi_n^{\circ 2}+\e_n^{\circ 2}+2\bpi_n\circ \e_n)^\top
    \end{pmatrix}=\H+\begin{pmatrix}
    \mathbb{E}(\e_1^{\circ 2})^\top\\
    \vdots\\
    \mathbb{E}(\e_n^{\circ 2})^\top
    \end{pmatrix}.
\end{aligned}    
\end{equation}
Since $\mathbb{E}(\e_i^{\circ 2})=\u_i/p$, we can get that
\begin{equation*}
    \B_2=\mathbb{E}(\hat{\H})-\H=\frac{1}{p}[\u_1,\ldots,\u_n]^\top.
\end{equation*}

\subsection{Significant IPA canonical pathways in Section \ref{sec::rosmap}}\label{sec::pathways}
We list all significant IPA canonical pathways from \texttt{bMIND} in Table~\ref{ipa1} and \texttt{bMIND+DECALS} in \ref{ipa2}. 
For each cell type, the pathways are listed by the increasing order of BH p-values. 
If there is no significant pathways in the cell type, it is left as ``NA'' in the table.
\begin{table}[!t]
{\small\centering
{\renewcommand{\arraystretch}{0.9}
\begin{tabular}{|c|l|}\hline
Cell Type & Pathways\\\hline
Neu & NA\\\hline
Oli & NA\\\hline
\multirow{3}{*}{Ast} & EIF2 Signaling,
Regulation of eIF4 and p70S6K Signaling,\\
& Inhibition of ARE-Mediated mRNA Degradation Pathway, Apelin Muscle Signaling Pathway,\\
& Huntington's Disease Signaling, 
Clathrin-mediated Endocytosis Signaling.\\ \hline
\multirow{8}{*}{Mic} & EIF2 Signaling,
CLEAR Signaling Pathway,
mTOR Signaling, 
Mitochondrial Dysfunction,\\
& Regulation of eIF4 and p70S6K Signaling,
Oxidative Phosphorylation,\\
& Protein Ubiquitination Pathway,
Assembly of RNA Polymerase II Complex,\\
& Androgen Signaling,
Epithelial Adherens Junction Signaling,
 Molecular Mechanisms of Cancer,\\
& Spliceosomal Cycle,
Oxytocin Signaling Pathway,\\
& Role of MAPK Signaling in Promoting the Pathogenesis of Influenza,\\
& Phagosome Maturation,
Estrogen Receptor Signaling,
Amyloid Processing,\\
& RANK Signaling in Osteoclasts,
Notch Signaling.\\ \hline
End & NA\\ \hline
\end{tabular}}}
\caption{Significant pathways identified from \texttt{bMIND}.}
\label{ipa1}
\end{table}

\begin{table}[!t]
{\small\centering
{\renewcommand{\arraystretch}{0.9}
\begin{tabular}{|c|l|}\hline
 Cell Type & Pathways\\\hline
Neu & Synaptogenesis Signaling Pathway.\\\hline
\multirow{4}{*}{Oli} & CSDE1 Signaling Pathway,
Mitochondrial L-carnitine Shuttle Pathway,\\
& LPS/IL-1 Mediated Inhibition of RXR Function,
Noradrenaline and Adrenaline Degradation,\\
& Aryl Hydrocarbon Receptor Signaling, 
L-DOPA Degradation,
Sumoylation Pathway.\\\hline
\multirow{2}{*}{Ast} & EIF2 Signaling,
Regulation of eIF4 and p70S6K Signaling,
mTOR Signaling,\\
& Coronavirus Pathogenesis Pathway,
Noradrenaline and Adrenaline Degradation.\\ \hline
\multirow{26}{*}{Mic} & EIF2 Signaling,
mTOR Signaling,
Regulation of eIF4 and p70S6K Signaling,\\
& Fatty Acid $\beta$-oxidation I,
Superpathway of Cholesterol Biosynthesis,\\
& Molecular Mechanisms of Cancer,
Cholesterol Biosynthesis I,\\
& Cholesterol Biosynthesis II (via 24,25-dihydrolanosterol),\\
&Cholesterol Biosynthesis III (via Desmosterol),\\
& Superpathway of D-myo-inositol (1,4,5)-trisphosphate Metabolism,\\
& 1D-myo-inositol Hexakisphosphate Biosynthesis II (Mammalian),\\
& Putrescine Degradation III,
Hereditary Breast Cancer Signaling,
Dopamine Degradation,\\
& Insulin Receptor Signaling,
G$\alpha$q Signaling,
AMPK Signaling,
Sumoylation Pathway,\\
& G Protein Signaling Mediated by Tubby,
Ethanol Degradation IV,\\
& D-myo-inositol (1,4,5)-trisphosphate Degradation,
3-phosphoinositide Degradation, \\
& Coronavirus Pathogenesis Pathway, 
Noradrenaline and Adrenaline Degradation,\\
& NRF2-mediated Oxidative Stress Response,
Superpathway of Inositol Phosphate Compounds,\\
& Chronic Myeloid Leukemia Signaling,
D-myo-inositol (1,3,4)-trisphosphate Biosynthesis,\\
& PTEN Signaling,
Aryl Hydrocarbon Receptor Signaling,
TWEAK Signaling,\\
& D-myo-inositol-5-phosphate Metabolism,
 Signaling by Rho Family GTPases,\\
& Acute Phase Response Signaling,
Xenobiotic Metabolism Signaling,\\
& RAR Activation,
LPS/IL-1 Mediated Inhibition of RXR Function,\\
& IL-8 Signaling,
D-myo-inositol (1,4,5,6)-Tetrakisphosphate Biosynthesis,\\
& D-myo-inositol (3,4,5,6)-tetrakisphosphate Biosynthesis,\\
& Xenobiotic Metabolism General Signaling Pathway,
PDGF Signaling, \\
& CLEAR Signaling Pathway,
Ethanol Degradation II,
Apoptosis Signaling,\\
& Apelin Muscle Signaling Pathway,
CXCR4 Signaling,
BER (Base Excision Repair) Pathway,\\
& RHOGDI Signaling,
Death Receptor Signaling,
Integrin Signaling,\\
& Induction of Apoptosis by HIV1,
Oxidative Ethanol Degradation III,\\
& Sirtuin Signaling Pathway,
Isoleucine Degradation I.\\ \hline
End & NA \\ \hline
\end{tabular}}}
\caption{Significant pathways identified from \texttt{bMIND+DECALS}.}
\label{ipa2}
\end{table}

\end{document}